\documentclass[pdflatex,sn-mathphys-num]{sn-jnl}

\usepackage{graphicx}%
\usepackage{multirow}%
\usepackage{amsmath,amssymb,amsfonts}%
\usepackage{amsthm}%
\usepackage{mathrsfs}%
\usepackage[title]{appendix}%
\usepackage{xcolor}%
\usepackage{textcomp}%
\usepackage{manyfoot}%
\usepackage{booktabs}%
\usepackage{algorithm}%
\usepackage{algorithmicx}%
\usepackage{algpseudocode}%
\usepackage{listings}%
\usepackage[numbers]{natbib}  
\usepackage{amsmath} 

\usepackage{float}  
\usepackage{tabularx}
\usepackage{array}

\theoremstyle{thmstyleone}%
\theoremstyle{thmstyletwo}%

\theoremstyle{thmstylethree}%

\begin{document}

\title[Article Title]{Conditional effects of cross-product substitution on systemic risk in multilayer food trade networks}


\author[2]{\fnm{Feiyan} \sur{Guo}}

\author[1]{\fnm{Jianlin} \sur{Zhou}}

\author[3]{\fnm{Lin} \sur{Qi}}

\author*[1]{\fnm{Ying} \sur{Fan}}\email{yfan@bnu.edu.cn}

\affil*[1]{\orgdiv{School of Systems Science}, \orgname{Beijing Normal University}, \orgaddress{\city{Beijing}, \postcode{100875}, \country{P. R. China}}}

\affil[2]{\orgdiv{School of National Safety and Emergency Management}, \orgname{Beijing Normal University}, \orgaddress{\city{Beijing}, \postcode{100875}, \country{P. R. China}}}

\affil[3]{\orgdiv{College of Management Science and Engineering}, \orgname{Beijing Information Science and Technology University}, \orgaddress{\city{Beijing}, \postcode{102206}, \country{P. R. China}}}

\abstract{Localized shocks arising from climate extremes, geopolitical conflicts, and trade protectionism cascade through trade networks, triggering global food crises. Cross-product substitution, a critical response strategy, induces cross-product cascading effects that remain underexplored. Here, we develop a multilayer network model that simulates the short-term response to food supply shocks. When applied to cereal trade networks, comparisons with and without substitution, as well as with increased substitute layers, reveal that substitution mitigates risks in the shocked layer but induces derived risks in substitute layers, causing the network system to present four response regimes ranging from resilient to systemic crisis. These regimes' boundaries and magnitudes emerge from the interplay of four critical factors: shock intensity, substitution extent, supply capacity of substitute layers, and inter-layer substitution structure. Scenario simulations of three real-world shocks further reveal country-level heterogeneity in substitution effectiveness. Our framework provides a quantitative tool for designing response strategies and resilient food systems.}

\keywords{food security, food trade, systemic risk, cross-product substitution, multilayer network}

\maketitle

\section{Introduction}\label{sec:intro}

Global food security faces multiple threats from climate extremes, geopolitical conflicts, and trade protectionism \cite{1,2,3,4,5,6,7}. While these localized shocks directly impair food supply in target countries, they also cascade through highly interconnected trade networks, rapidly propagating to distant economies and potentially culminating in a global supply crisis \cite{8,9,10,11,12,13,14}. Scientifically assessing these cascading risks and designing effective response strategies is thus essential for enhancing food system resilience and advancing the Sustainable Development Goal of ``Zero Hunger''.

Paulus et al. systematically evaluated five modeling approaches for global food system risk assessment \cite{15}: equilibrium-based \cite{16}, input-output \cite{17}, network \cite{18}, agent-based \cite{19}, and system dynamics models \cite{20}. Among these, network modeling offers distinct advantages in analyzing structural cascading effects \cite{15}. By characterizing trade topology with real flow data \cite{21,22,23}, it effectively captures shock-response dynamics and identifies risk-propagation pathways \cite{24,25,26,27,28}. Such models have been used to assess the cascading impacts of localized shocks \cite{9,18,29,30,31} and quantify the effects of country-level response strategies, including reserve release \cite{26,32} and trade adjustments \cite{25,26,27,28,33}, on risk mitigation and propagation.

Cross-product substitution (e.g., wheat for rice) also represents a critical country-level response strategy for food supply shocks, yet its induced cross-product cascading effects remain underexplored \cite{9,27,34}. Existing single-layer network models can only simulate shock propagation within a product's own trade network, failing to capture cross-network cascading feedback and rendering them unsuitable for quantifying substitution effects \cite{26,27,28}. While there is a multi-layer network model incorporating cross-product substitution strategy, it focuses on medium- to long-term adaptive adjustments, making it difficult to reflect response characteristics under a short-term out-of-equilibrium condition induced by supply shocks \cite{35}.

Here, we present a multilayer network-based shock response model incorporating cross-product substitution to investigate the short-term responses and their cascading effects of sudden supply shocks originating from a single product. The model abstracts trade networks of substitutable products as interconnected network layers and, by introducing country-level trade adjustment and cross-product substitution strategies, enables bidirectional dynamic simulation of intralayer shock propagation and cross-layer feedback loops (see Section~\ref{model} for details). Applying the model to cereals as a major component of global food trade and using their annual production, reserve, and trade data, we conduct three analyses: (1) quantifying the effects of cross-product substitution on network-level supply risk following supply shocks; (2) revealing how increased product layers influence cross-crop substitution effectiveness; and (3) evaluating the country-level effectiveness of cross-product substitution and identifying root causes of national supply-shortage risks. With this work, we establish a quantitative tool for designing response strategies and resilient food systems.

\section{Results}\label{sec:result}

\subsection{Effects of cross-product substitution on network supply risk}\label{sub:effect}
\begin{figure}[h]
\centering
\includegraphics[width=0.9\textwidth]{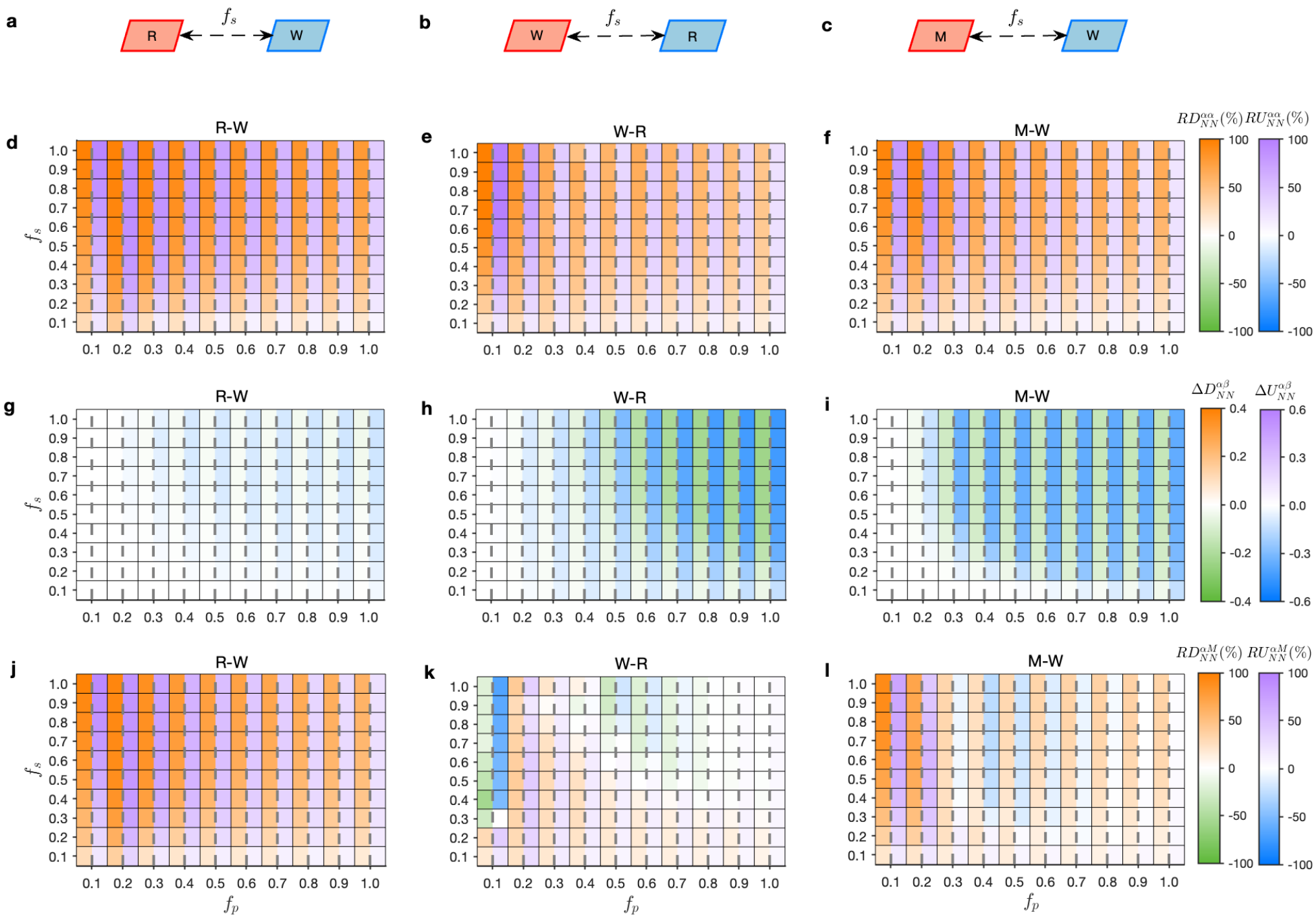}
\caption{Effects of two-layer substitution relative to no-substitution. \textbf{a-c}, Schematic topology of simulation scenarios. Red quadrilaterals denote the shocked layer, blue quadrilaterals denote the substitute layer, with bidirectional dashed lines between layers indicating substitution relationships; $f_s$ denotes the substitution fraction. Product layers are abbreviated as R = rice, W = wheat, and M = maize, constituting three scenarios: R-W \textbf{(a)}, W-R \textbf{(b)}, and M-W \textbf{(c)}.  \textbf{d-l}, Under shock intensity $f_p \in[0.1,1.0]$ (step size 0.1), effects of two-layer substitution ($f_s \in[0.1,1.0]$, step size 0.1) relative to no-substitution ($f_s=0.0$) on the shocked layer \textbf{(d-f)}, the substitute layer \textbf{(g-i)}, and the entire two-layer network  \textbf{(j-l)}. Each cell displays the effect for the fixed ($f_p$, $f_s$) combination and is split into two halves by a gray dashed line. In \textbf{(d-f)} and \textbf{(j-l)}, the left half shows deficit compensation rate ($R D_{N N}^{\alpha \alpha}$, $R D_{N N}^{\alpha M}$), while the right half shows deficit unevenness reduction rate ($R U_{N N}^{\alpha \alpha}$, $R U_{N N}^{\alpha M}$). In \textbf{(g-i)}, the left half indicates differential deficit ($\Delta D_{N N}^{\alpha \beta}$), and the right half shows differential deficit unevenness ($\Delta U_{N N}^{\alpha \beta}$).}
\label{fig1}
\end{figure}

Based on two-layer trade networks composed of shocked products and their strong substitutes, the cascading effects of cross-product substitution can be quantified by comparing shock impacts (consumption deficit) between no-substitution and two-layer substitution simulations. Figure \ref{fig1}a-c illustrates three two-layer network simulation scenarios formed by the three staples (rice, wheat, and maize) and their strong substitutes \cite{34,35}: (1) rice shocked with wheat substitution (R-W); (2) wheat shocked with rice substitution (W-R); and (3) maize shocked with wheat substitution (M-W).

\bigskip
Figure \ref{fig1}d--l presents the effects of two-layer substitution relative to no-substitution baseline across three scenarios. For the shocked layer (Fig. \ref{fig1}d--f) and overall two-layer network (Fig. \ref{fig1}j--l), cells show relative changes in shock impact, represented as deficit compensation rate ($R D>0$: consumption deficit compensation; $R D<0$: amplification) and deficit unevenness reduction rate ($R U>0$: impact inequality reduction, $R U<0$: aggravation). For the substitute layer (Fig. \ref{fig1}g--i), where the no-substitution baseline is zero, cells instead display absolute changes, represented as differential deficit ($\Delta D<0$: deficit amplification) and differential deficit unevenness ($\Delta U<0$: inequality aggravation). The effects on shocked and substitute layers exhibit cross-scenario consistency: while cross-product substitution mitigates shock impacts ($R D_{N N}^{\alpha \alpha}>0$, $R U_{N N}^{\alpha \alpha}>0$) in the shocked layer, it induces derived impacts ($\Delta D_{N N}^{\alpha \beta}<0$, $\Delta U_{N N}^{\alpha \beta}<0$) in the substitute layer. The overall network displays four response regimes:

(i) Efficient compensation with inequality reduction ($R D_{N N}^{\alpha M}>0$, $R U_{N N}^{\alpha M}>0$): Both average deficit and deficit unevenness declined, representing efficient risk mitigation.

(ii) Efficient compensation with inequality aggravation ($R D_{N N}^{\alpha M}>0$, $R U_{N N}^{\alpha M}<0$): Deficits are compensated on average yet unevenly distributed, concentrating vulnerability in certain countries.

(iii) Deficit amplification with inequality aggravation ($R D_{N N}^{\alpha M}<0$, $R U_{N N}^{\alpha M}<0$): Both average deficit and deficit unevenness worsen, signaling cascading crisis;

(iv) Deficit amplification with inequality reduction ($R D_{N N}^{\alpha M}<0$, $R U_{N N}^{\alpha M}>0$): Average deficits worsen while unevenness narrows, characteristic of systemic collapse.

Regimes (i)–(ii) represent resilient responses, while (iii)–(iv) correspond to systemic crisis states.

\bigskip
Comparing Fig. \ref{fig1}j--l reveals significant scenario heterogeneity in the overall network response. The R-W scenario performs optimally (Fig. \ref{fig1}j), delivering effective compensation and inequality reduction across the entire parameter space, with the highest mitigation efficiency for the shocked layer (Fig. \ref{fig1}d) and minimal side effects on the substitute layer (Fig. \ref{fig1}g). This may be attributable to the wheat layer's robust supply capacity, characterized by abundant supply, high efficiency, and stability, which enables effective shock absorption with minimal internal deficits. Network supply properties are provided in Table \ref{tab2}.

\bigskip
The M-W scenario ranks second (Fig. \ref{fig1}l), achieving effective compensation across the entire parameter space but exhibiting aggravated inequality within the region $f_p \in[0.3,0.8]$, $f_s \in[0.3,1.0]$. This phenomenon likely stems from the maize layer's insufficient trade path diversification and high export concentration, whereby cross-layer feedback loops synchronize the amplification of deficit in import-dependent countries across both layers, thereby exacerbating impact inequality.

The W-R scenario performs worst (Fig. \ref{fig1}k), with both deficits and inequality worsening in the region $f_p \in[0.5,0.8]$, $f_s \in[0.5,1.0]$ and precipitating systemic collapse when $f_p \in[0.9,1.0]$, and $f_s \in[0.4,1.0]$. This likely originates from the rice layer's weak supply capacity, marked by insufficient supply efficiency and excessive export market concentration, which constrains its ability to absorb shocks effectively and ultimately amplifies systemic impact.

\bigskip
Cross-product substitution effectiveness exhibits strong parametric sensitivity to shock intensity $f_p$ and substitution fraction $f_s$. This sensitivity displays non-monotonic behavior with optimal intervals (Fig. \ref{fig1}d--l). At fixed $f_p$, increasing $f_s$ enhances the compensation rate for the shocked layer and amplifies deficits in the substitute layer, yet with a diminishing rate. At certain $f_p$ levels, an inflection point even emerges when $f_s\geq0.8$, beyond which both compensation rate and substitute-layer deficit begin to decline. This suggests cross-product substitution operates within an optimal range, and over-substitution erodes its effectiveness. For the overall network, increasing $f_s$ yields scenario-dependent effects on compensation rate: monotonic increase in the R-W scenario (Fig. \ref{fig1}j), pronounced fluctuations in the W-R scenario (Fig. \ref{fig1}k) with optimal $f_s \in[0.2,0.5]$, and both trends coexist in the M-W scenario (Fig. \ref{fig1}l).

At fixed $f_s$, the compensation rate for the shocked layer generally increases first and then decreases as $f_p$ increases, peaking at $f_p=0.2$ rather than following the intuitively expected monotonic decline. This counterintuitive pattern also manifests in the overall network, where compensation rate fluctuates with $f_p$ but peaks predominantly at $f_p=0.2$. Particularly in the W-R scenario with $f_s \in[0.3,1.0]$, deficit amplification with inequality aggravation occurs under low shock intensity ($f_p=0.1$) (Fig. \ref{fig1}k). This indicates that moderate shocks more effectively activate substitution responses. Conversely, deficits on the substitute layer increase monotonically with $f_p$, reflecting the logical positive correlation between shock magnitude and shock dispersion.

\subsection{Effects of increased product layers on substitution effectiveness}\label{sub:added}

\begin{figure}[ht]
\centering
\includegraphics[width=0.9\textwidth]{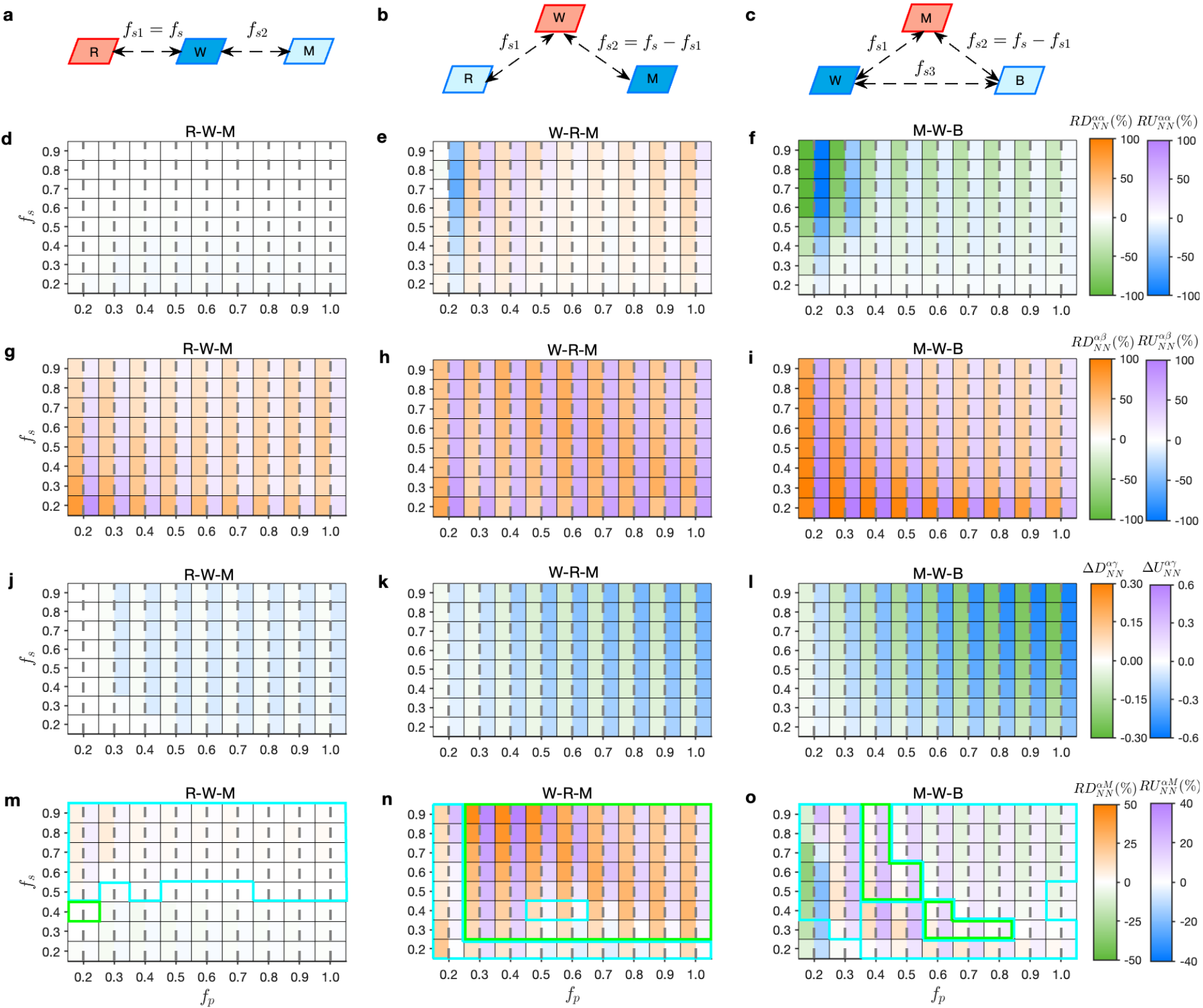}
\caption{Effects of three-layer substitution relative to the two-layer substitution. \textbf{a-c}, Schematic topology of simulation scenarios. Red quadrilaterals denote the shocked layer, blue quadrilaterals denote the substitute layer (dark blue = strong supply capacity, light blue = weak). Bidirectional dashed lines indicate substitution relationships, where $f_s$ represents the total substitution fraction for shocked layer distributed across $f_{s1}$, $f_{s2}$, and $f_{s3}$. Add product layer abbreviation B = barley yields three scenarios: R-W-M chain structure \textbf{(a)}, W-R-M triadic structure \textbf{(b)}, and M-W-B triangular structure \textbf{(c)}. \textbf{d-o}, Under $f_p \in[0.2,1.0]$ and $f_s \in[0.2,0.9]$ (both with step size 0.1), effects of three-layer substitution (non-zero ($f_{s1}$, $f_{s2}$) or ($f_{s1}$, $f_{s2}$, $f_{s3}$) combinations) relative to two-layer substitution ($f_{s1} = f_s$, $f_{s2} = f_{s3} = 0$) on shocked layer \textbf{(d-f)}, first substitute layer (the substitution fraction with the shocked layer is $f_{s1}$) \textbf{(g-i)}, second substitute layer \textbf{(j-l)}, and entire three-layer network \textbf{(m-o)}. Each cell represents the mean effects of all non-zero  ($f_{s1}$, $f_{s2}$)  or ($f_{s1}$, $f_{s2}$, $f_{s3}$)  combinations (step size 0.1) at fixed ($f_p$, $f_s$) and is split into two halves by a gray dashed line. In \textbf{(d-f)}, \textbf{(g-i)}, and \textbf{(m-o)}, the left half shows deficit compensation rate ($R D_{N N}^{\alpha \alpha}$, $R D_{N N}^{\alpha \beta}$, $R D_{N N}^{\alpha M}$), while the right half shows deficit unevenness reduction rate ($R U_{N N}^{\alpha \alpha}$, $R U_{N N}^{\alpha \beta}$, $R U_{N N}^{\alpha M}$). In \textbf{(j-l)}, the left half indicates differential deficit ($\Delta D_{N N}^{\alpha \gamma}$), and the right half shows differential deficit unevenness ($\Delta U_{N N}^{\alpha \gamma}$). In \textbf{(m-o)}, blue boxes indicate cells where optimal non-zero ($f_{s1}$, $f_{s2}$)  or ($f_{s1}$, $f_{s2}$, $f_{s3}$) combinations exist that maximize the compensation rate with $f_{s1} \geq f_{s2}$; green boxes indicate such combinations where $f_{s1} < f_{s2}$. Cells outside these boxes indicate that all possible combinations reduce the compensation rate, meaning no optimal combination exists.}
\label{fig2}
\end{figure}

Increasing product layers may reconfigure shock propagation pathways, thereby influencing cross-product substitution effectiveness. To assess this, we compare shock impacts between two-layer substitution and three-layer substitution simulations. The two-layer networks in Fig. \ref{fig1}a--c are extended to three-layer simulation scenarios (Fig. \ref{fig2}a--c) based on product substitutability \cite{35}: (1) rice shocked with wheat and maize substitution (R-W-M chain inter-layer substitution structure); (2) wheat shocked with rice and maize substitution (W-R-M triadic structure); and (3) maize shocked with wheat and barley substitution (M-W-B triangular structure). Across these three scenarios, the effects of three-layer substitution relative to no-substitution (Fig. \ref{fig7}) exhibit consistency with two-layer network conclusions, corroborating the robustness of cross-product substitution effects.

\bigskip
Fig. \ref{fig2}d--o illustrates the effects of three-layer substitution relative to two-layer substitution. The core effect of increased substitute product layers is characterized by feedback loops and multi-path shock dispersion. Three-layer substitution attenuates the mitigation efficiency for the shocked layer via cross-layer feedback loops (Fig. \ref{fig2}d--f), particularly pronounced in the R-W-M chain and M-W-B triangular structures. Meanwhile, the added second substitute layer mitigates shock impacts on the existing first substitute layer (Fig. \ref{fig2}g--i) while generating derived impacts within itself (Fig. \ref{fig2}j--l). This demonstrates that added substitute layers divert substitution demand from the original substitute layer to the new layer, achieving shock dispersion.

The overall network displays four response regimes (Fig. \ref{fig2}m--o) consistent with Fig. \ref{fig1}j--l, showing pronounced scenario heterogeneity. The R-W-M chain structure performs poorest (Fig. \ref{fig2}m), constrained by its serial substitution pathway that precludes parallel shock dispersal. The added second substitute layer (maize) can only receive residual substitution demand from the first substitute layer ($f_{s2} \in [0.1, 1 - f_{s1}$]). When $f_{s1} = f_s \in [0.2, 0.4]$, although $f_{s2}$ retains operational flexibility, the maize layer's inferior supply capacity relative to the wheat layer acts as a bottleneck, dragging down the compensation rate. When $f_{s1} = f_s \in [0.5, 0.9]$, while optimal ($f_{s1}$, $f_{s2}$) combinations (blue boxes) exist that can enhance the compensation rate, the absence of the parallel architecture renders efficiency gains substantially inferior to triadic and triangular structures.

The W-R-M triadic structure delivers optimal performance (Fig. \ref{fig2}n), benefiting from its parallel substitution characteristic. Both substitute layers simultaneously receive substitution demand from the shocked layer ($f_s = f_{s1} + f_{s2}$), achieving synergistic shock dispersal. Optimal ($f_{s1}$, $f_{s2}$) combinations (green boxes) predominantly exhibit a ``strong-layer-dominant, weak-layer-auxiliary'' configuration ($f_{s1}<f_{s2}$). This configuration avoids overloading the weak layer (rice) while fully unleashing the strong layer's (maize) potential, enabling the system to maximize utilization of both layers' substitution capacities and achieve the most significant efficiency gains.

The M-W-B triangular structure exhibits performance fluctuations (Fig. \ref{fig2}o), attributable to its complex triangular loops. Beyond parallel substitution ($f_s = f_{s1} + f_{s2}$), the incorporation of $f_{s3} \in \left[0.1, \min(1 - f_{s1}, 1 - f_{s2})\right]$ regenerates cross-layer feedback, forming a tridirectionally cyclic substitution. This may either enhance the compensation rate via shock dispersion or attenuate it through feedback loops. Most optimal ($f_{s1}$, $f_{s2}$, $f_{s3}$) combinations (blue boxes) require limiting the weak layer's share (lower $f_{s2}$ with barley), activating the strong layer (higher $f_{s1}$ with wheat), and suppressing the interference of $f_{s3}$ to achieve stable efficiency gains within complex loops.

\subsection{Country-level effectiveness of cross-product substitution}\label{sub:country}

To validate country-level substitution effectiveness, we conduct scenario simulations based on three representative shocks: (1) Indian rice export ban: July 20, 2023 total prohibition of non-basmati broken rice exports \cite{36}; (2) Ukrainian wheat export disruption: Russia--Ukraine conflict caused a 39\% decline in exports between 2021 and 2022 \cite{32}; and (3) U.S. maize production shortfall: extreme heat and drought reduced 2012 output by 21\% compared to the preceding five non-drought years \cite{37}. For each scenario, we employ the corresponding two-layer network in Fig. \ref{fig1}a--c to conduct both the no-substitution and with-substitution simulations.

\begin{figure}[ht]
\centering
\includegraphics[width=0.82\textwidth]{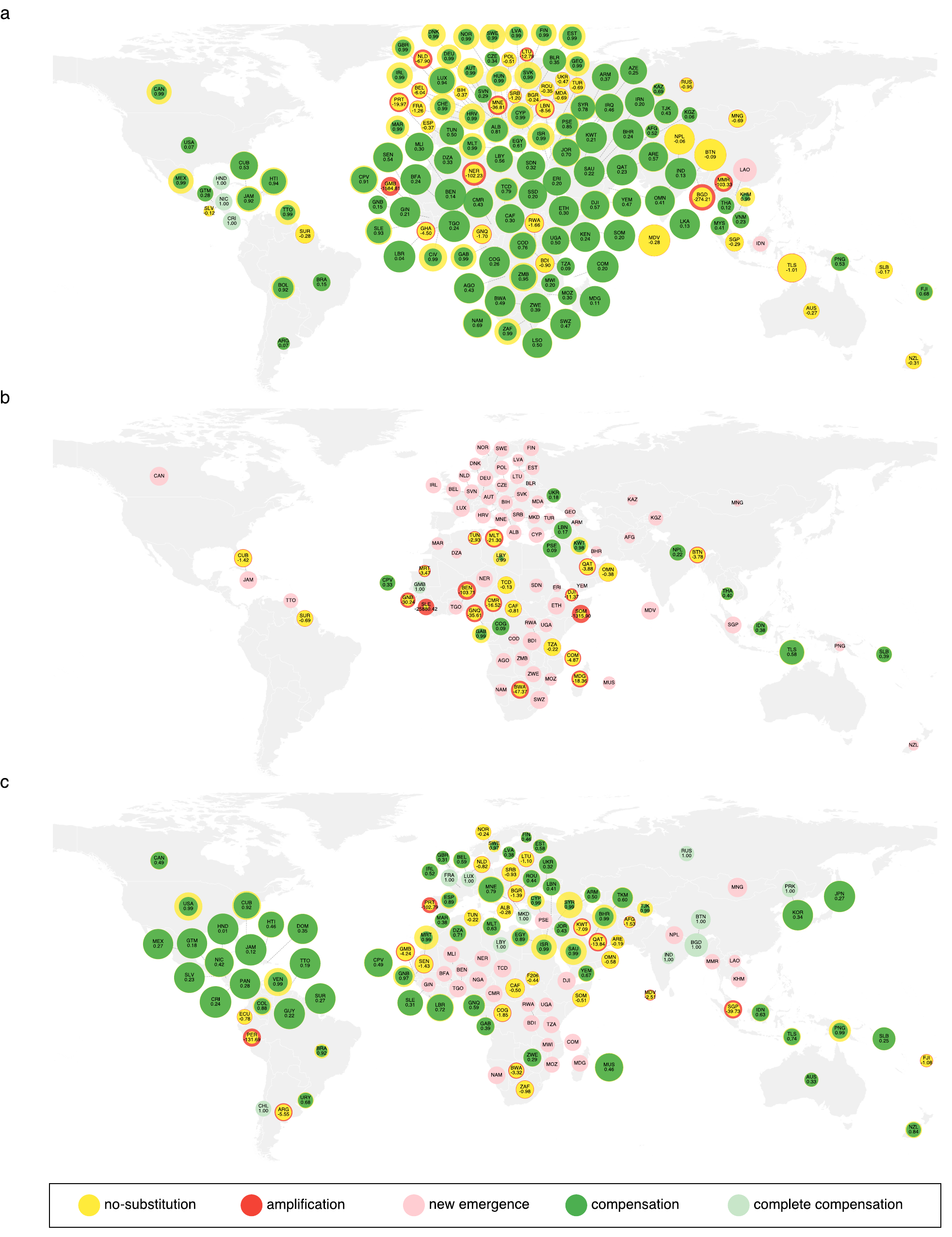}
\caption{Country-Level effects of two-layer substitution relative to no-substitution. Consumption deficits ($D_{i k}^{\alpha M}$) on countries under no-substitution ($f_s=0$) and two-layer substitution ($f_s=0.2$) for India's complete rice production halt ($f_p=1.0$) \textbf{(a)}, Ukraine's 40\% wheat production reduction ($f_p=0.4$) \textbf{(b)}, and the U.S. 21\% maize production reduction ($f_p=0.21$) \textbf{(c)}. Compared to deficits under no-substitution (yellow circles), some countries achieve deficit compensation (green circles) or even complete compensation (light green circles) through cross-product substitution, while others face deficit amplification (red circles) or new deficit emergence (light red circles). Circle sizes correspond to consumption deficits on a logarithmic scale, and values inside circles represent the deficit compensation rate ($R D_{i k}^{\alpha M}$, positive values indicate effective compensation, negative values indicate deficit amplification). Base map: World Bank Official Boundaries (CC BY 4.0), 2025 version, https://datacatalog.worldbank.org/search/dataset/0038272. In \textbf{(c)}, F206 (Sudan (former)) was generated by merging modern SDN (Sudan) and SSD (South Sudan) boundaries to represent the unified Sudan prior to the 2011 secession.}
\label{fig3}
\end{figure}

\bigskip
\textbf{Indian Rice Scenario:} Figure \ref{fig3}a reveals heterogeneous effects of cross-product substitution on countries. Relative to no-substitution (yellow circles), some countries achieve deficit compensation (green circles) or even complete compensation (light green circles) through substitution responses by themselves or other countries, while others face deficit amplification (red circles) or new deficit emergence (light red circles).

Countries severely impacted under no-substitution, notably the United Arab Emirates and Oman, benefit substantially from wheat substitution. Conversely, many countries, including Maldives and Niger, experience amplified deficits. Both the UAE and the Maldives exhibit extreme import dependence on Indian rice ($\geq 80\%$ of supply). The UAE leverages adequate wheat reserves and diversified import sources (e.g., major wheat exporters like Russia) to effectively compensate deficits, whereas the Maldives suffers exacerbated deficits due to wheat reserve inadequacy and similarly concentrated dependence on India. Similarly, Niger, despite low Indian rice dependence and thus a small baseline deficit, exhibits significant deficit amplification due to feedback loops, given its high import dependence for both rice and wheat. In contrast, countries like Laos, which are largely self-sufficient in rice but highly dependent on wheat imports, experience clear deficit emergence, as substitution directly exposes them to derived shocks.

Severely affected countries are concentrated in Africa and Asia, while European countries achieve more pronounced compensation. Network-wide, wheat substitution compensates deficits in over two-thirds of countries, achieving a compensation rate of 23.06\% and reducing inequality by 12.10\%, confirming its effectiveness.

Consistent with adaptive adjustment outcomes in ref. \cite{35}, our simulations confirm that countries like the UAE can compensate deficits while the Maldives face amplification, validating the robustness of heterogeneous responses across these countries. However, Niger and other import-dependent countries experience deficit amplification, contrasting with net rice gains reported in ref. \cite{35}. This discrepancy highlights that short-term responses may expose vulnerabilities undetectable in medium- to long-term adaptive scenarios.

\bigskip
\textbf{Ukrainian Wheat Scenario:} This shock has a limited impact on most countries, yet rice substitution still induces heterogeneous responses (Fig. \ref{fig3}b). Under no-substitution, Timor-Leste experiences the most severe impacts from indirect trade shocks. Although it does not directly import wheat from Ukraine, it is highly dependent on Indonesia, which decreased exports after its own direct trade with Ukraine was disrupted. Under rice substitution, Timor-Leste and Indonesia can redirect rice imports to India and Vietnam, achieving deficit compensation. However, countries with high import dependence on both wheat and rice, such as Sierra Leone, and countries like Canada that depend heavily on rice imports with high concentration, face significant deficit amplification or emergence.

Shock impacts show significant regional differentiation, with African countries most severely affected and exhibiting the most prominent deficit amplification or emergence. Despite derived shocks causing new deficits in many countries, the network-wide deficit compensation rate reaches 54.40\% and inequality is reduced by 58.50\%, confirming that cross-product substitution achieves effective mitigation through shock dispersion.

Compared to ref. \cite{32}, which only focuses on countries directly dependent on Ukrainian wheat, our results capture the significant impacts of cross-product cascades on indirectly linked countries like Timor-Leste. This underscores the necessity of accounting for cascading risks in risk assessment.

\bigskip
\textbf{U.S. Maize Scenario:} Despite weaker shock intensity than previous scenarios, the U.S. maize shock still generates substantial impacts on select countries (Fig. \ref{fig3}c). Countries severely affected under no-substitution, notably Costa Rica and Jamaica, receive partial compensation under wheat substitution but remain among the most impacted. These countries are highly import dependent on the U.S. for maize and wheat, making them significantly vulnerable once the U.S. suffers a shock. By contrast, countries like Cuba and Venezuela substantially mitigate impacts by redirecting imports to major wheat exporters such as France and China. For countries experiencing deficit amplification or emergence, negative substitution effects are limited, as their baseline impacts were small.

The Americas constitute the disaster epicenter, while wheat substitution delivers more pronounced compensation in Asia and Europe. While increasing substitute layers could enhance compensation efficiency in heavily affected regions, we limit the analysis to two-layer substitution, which already achieves a network-wide compensation rate of 25.44\% and inequality reduction of 21.10\%, delivering macro-level effective shock mitigation.

\section{Discussion}\label{sec:discu}

We propose a shock response model incorporating cross-product substitution, complementing existing network modeling aimed at assessing cascading risks in global food systems. The model constructs a multilayer food trade network accounting for trade dependencies and product substitutability, and enables bidirectional shock propagation simulation within and across layers by introducing country-level trade adjustment and cross-product substitution strategies. This framework dissects short-term, nonlinear and out-of-equilibrium responses of multilayer trade networks to supply shocks, offering tools for response strategy formulation and resilient food system design.

First, we reveal that while cross-product substitution can mitigate risks in shocked layers, it generates derived risks in substitute layers, causing the overall network to present four response regimes ranging from resilient to systemic crisis. These regime transitions depend on the triple interaction among shock intensity, substitution fraction, and substitute-layer supply capacity. Moderate shocks effectively activate cross-product substitution responses, whereas overly weak shocks suffer from insufficient activation motives, yielding low efficiency and even concentrating vulnerability to amplify risk. While increasing substitution fractions enhances mitigation efficiency, supply constraints in substitute layers cause diminishing mitigation magnitude, with exceeding critical thresholds triggering risk amplification. Scenario heterogeneity reveals that products with strong supply capacity (e.g., wheat) as substitute layers achieve optimal risk mitigation, while weak-capacity layers (e.g., rice) readily trigger amplification effects. Therefore, effective shock response requires concurrent differentiated configuration and dynamic optimization, prioritizing strong substitute products and adjusting substitution fractions dynamically based on shock intensity to achieve resilient outcomes. 

Second, we identify multi-path shock dispersion facilitated by increased product layers, whose efficiency in enhancing cross-product substitution effectiveness depends on the coupling between inter-layer substitution structure and substitution fractions. The added substitute layer's value is diverting substitution demand to disperse risk, while the costs include its own derivative risks and the weakening of the shocked layer's mitigation efficiency caused by feedback loops. In three-layer trade networks formed by adding one product layer, the chain inter-layer substitution structure suffers from low efficiency due to serial pathway characteristics and is easily dragged down by added weak substitute layers. The triangular structure's complex loops make its efficiency highly dependent on substitution fraction configurations. The triadic structure performs optimally by leveraging synergistic risk dispersal from parallel substitution. This structure employs a ``strong-layer-dominant, weak-layer-auxiliary'' substitution fraction configuration logic, achieving supply-capacity-weighted optimal task allocation. Therefore, implementing cross-product substitution strategies should encourage parallel substitution and prioritize activating strong-substitute products (e.g., wheat, maize) to bear primary substitution pressure, avoiding efficiency losses in systemic risk mitigation caused by equal pressure distribution.

Finally, scenario analyses expose significant heterogeneity in cross-product substitution effectiveness across countries, wherein substitution buffers shortages in some countries while amplifying or triggering novel supply crises in others. Focusing on characteristics of adversely affected countries reveals that commonalities include insufficient substitute product reserves, high import dependence, and high import concentration, which render substitute products themselves secondary shock sources. Conversely, beneficiary countries typically possess advantages such as adequate reserves, high self-sufficiency, or import source diversification, enabling effective absorption of substitution demand and risk dispersal. Therefore, the key to enhancing country resilience lies in increasing reserve levels and self-sufficiency, while advancing dual trade diversification in import sources and product varieties, prioritizing trade relationships with diversified exporters holding abundant reserves, thereby effectively leveraging product substitution's mitigation efficacy and avoiding impacts from shocked products.

The model has two primary limitations. First, it adopts a parsimonious specification to describe agent (national economies) behavior: limited substitution capacity and shock transmitted to trade partners are allocated proportionally, and simulations assume homogeneous reserve release and substitution fractions across countries. This simplification is primarily constrained by the aggregated nature of available production, reserve, and trade data, which precludes more refined model design and parameter estimation \cite{26}. Second, the response mechanism is modeled without prices and markets, neglecting dynamic feedback from competitive price adjustments triggered by supply-demand imbalances on product substitution and trade \cite{34,38}, potentially undermining the model's explanatory power for real-world market adjustment processes.

Despite these limitations, this study enhances understanding of cascading effects induced by cross-product substitution and deepens insights into multidimensional systemic risks inherent in food trade and substitution. The effectiveness of cross-product substitution in mitigating systemic risks emerges from the interplay of four critical factors: shock intensity, substitution extent (encompassing both the number of substitute product layers and substitution fraction configurations), supply capacity of substitute layers, and inter-layer substitution structure. The model framework is not limited to single-product production shocks but can be extended to scenarios including export disruptions and multi-product compound shocks. Moreover, its applicability can also be extended to other resource trade systems with substitute products, such as the global energy trade network disrupted by the Russia-Ukraine conflict, where a fossil-to-renewable energy transition occurs \cite{39,40}. Future research should focus on incorporating finer-grained national heterogeneity (e.g., dietary cultures, purchasing power) and market dynamic mechanisms to enhance the model's real-world explanatory power and predictive accuracy.

\section{Methods}\label{sec:methods}

\subsection{Modeling trade and substitution as a multilayer network}\label{sub:net}
\begin{figure}[ht]
\centering
\includegraphics[width=0.9\textwidth]{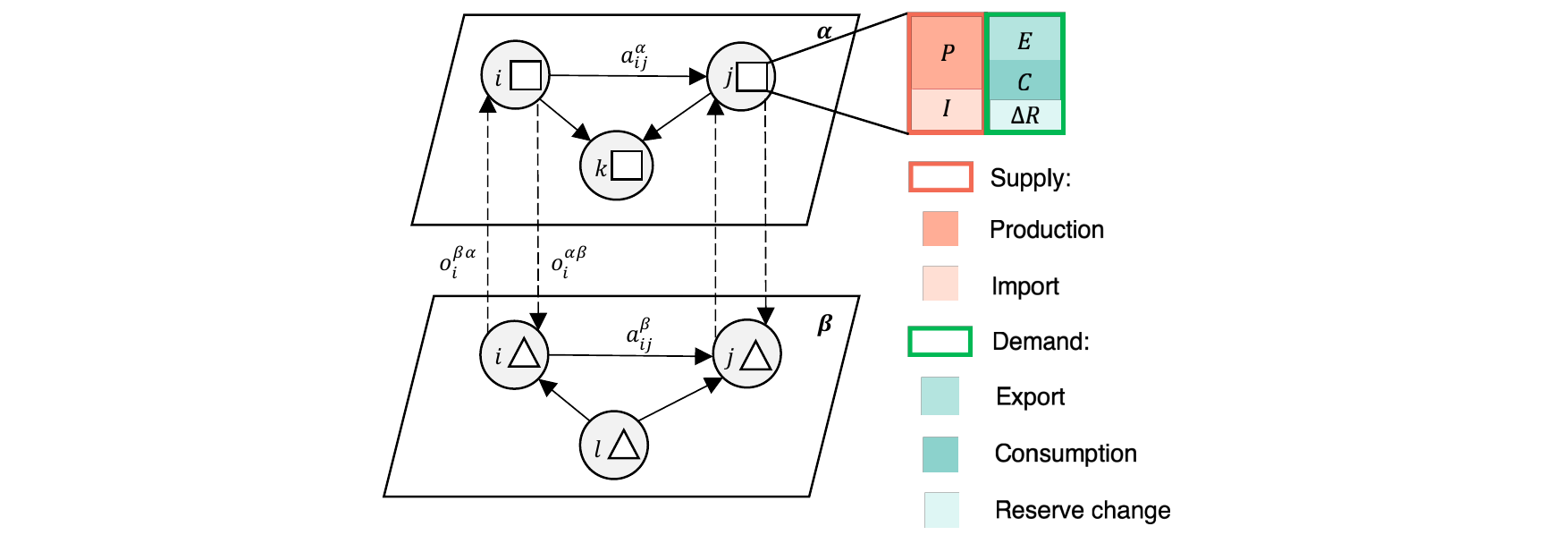}
\caption{Schematic representation of trade and substitution as a multilayer network for four countries and two products. Products (upper layer, square, $\alpha$; lower layer, triangle, $\beta$) for each country (gray circles) satisfy mass balance. Specifically, in the example square, the supply in the left half (the sum of production and imports, red) equals the demand in the right half (the sum of exports, consumption, and changes in reserves, green). Trade is described by weighted directed links (solid arrows) within each layer, where $a_{i j}^{\alpha}$ represents the volume of product $\alpha$ exported from country $i$ to country $j$. Substitution involves using products from other layers within a country to replace products in the current layer, thereby connecting different layers (dashed arrows).  $o_{i}^{\beta \alpha}$ denotes the substitution volume of product $\beta$ for product $\alpha$ in country $i$, where country $i$ acts as an in-neighbor on layer $\alpha$ and an out-neighbor on layer $\beta$ for this process. Conversely, $o_{i}^{\alpha \beta}$ represents the substitution volume of product $\alpha$ for product $\beta$ in country $i$.}
\label{fig4}
\end{figure}

Figure \ref{fig4} represents a simplified version of our multilayer network coupling global food trade and substitution. A country $i$ with a certain supply of product $\alpha$ acts as a node in the multilayer network. This supply consists of production $P_{i}^{\alpha}$ and imports $I_{i}^{\alpha}$, and is allocated according to demand into exports $E_{i}^{\alpha}$, consumption $C_{i}^{\alpha}$, and changes in reserves $\Delta R_{i}^{\alpha}$ (positive values indicate an increase in reserves), satisfying the mass balance at the country level: $P_i^\alpha + I_i^\alpha = E_i^\alpha + C_i^\alpha + \Delta R_i^\alpha$ (all in kcal equivalents) \cite{26}.

Inter-country trade of a specific product constitutes a single trade network layer. In Fig. \ref{fig4}, country $i$ exports product $\alpha$ to country $j$, forming a directed trade link from exporter $i$ to importer $j$, with the weight being the export volume $a_{i j}^{\alpha}$. All such links constitute a weighted directed trade matrix $A^{\alpha}$ describing product $\alpha$.

Intra-country substitution between two products constitutes a single inter-layer substitution network. In Fig. \ref{fig4}, products $\alpha$ and $\beta$  exhibit high substitutability. When country $i$ experiences insufficient supply of product $\alpha$, it can increase domestic demand for product $\beta$ as a substitute, forming a directed substitution link from country $i$ in layer $\beta$ to its replica in layer $\alpha$, with the weight being the substitution volume $o_{i}^{\beta \alpha}$; conversely, a substitution link with weight $o_{i}^{\alpha \beta}$ is formed. All such links constitute a weighted directed inter-layer substitution matrix $O^{\alpha\beta}$ describing products $\alpha$ and $\beta$. 

A multilayer trade network encompassing $M$ types of products can be represented as a supra-adjacency matrix $G=(A,O)$, as shown in equation  \ref{eq1}, where $A = \left\{A^{\alpha_1}, A^{\alpha_2}, \cdots, A^{\alpha_M}\right\}$ and $O = \left\{O^{\alpha_1 \alpha_2}, O^{\alpha_1 \alpha_3}, \cdots, O^{\alpha_n \alpha_m} \mid \alpha_n \neq \alpha_m \right\}{(\alpha_n, \alpha_m \in \{\alpha_1, \alpha_2, \cdots, \alpha_M\})}$.

\begin{equation}
G=\left[\begin{array}{cccc}A^{\alpha_1} & O^{\alpha_1 \alpha_2} & \cdots & O^{\alpha_1 \alpha_M} \\ O^{\alpha_2 \alpha_1} & A^{\alpha_2} & \cdots & O^{\alpha_2 \alpha_M} \\ \vdots & \vdots & \ddots & \vdots \\ O^{\alpha_M \alpha_1} & O^{\alpha_M \alpha_2} & \cdots & A^{\alpha_M}\end{array}\right].
\label{eq1}
\end{equation}

\subsection{Simulating iterative propagation of supply shocks}\label{sub:model}

\begin{figure}[ht]
\centering
\includegraphics[width=0.89\textwidth]{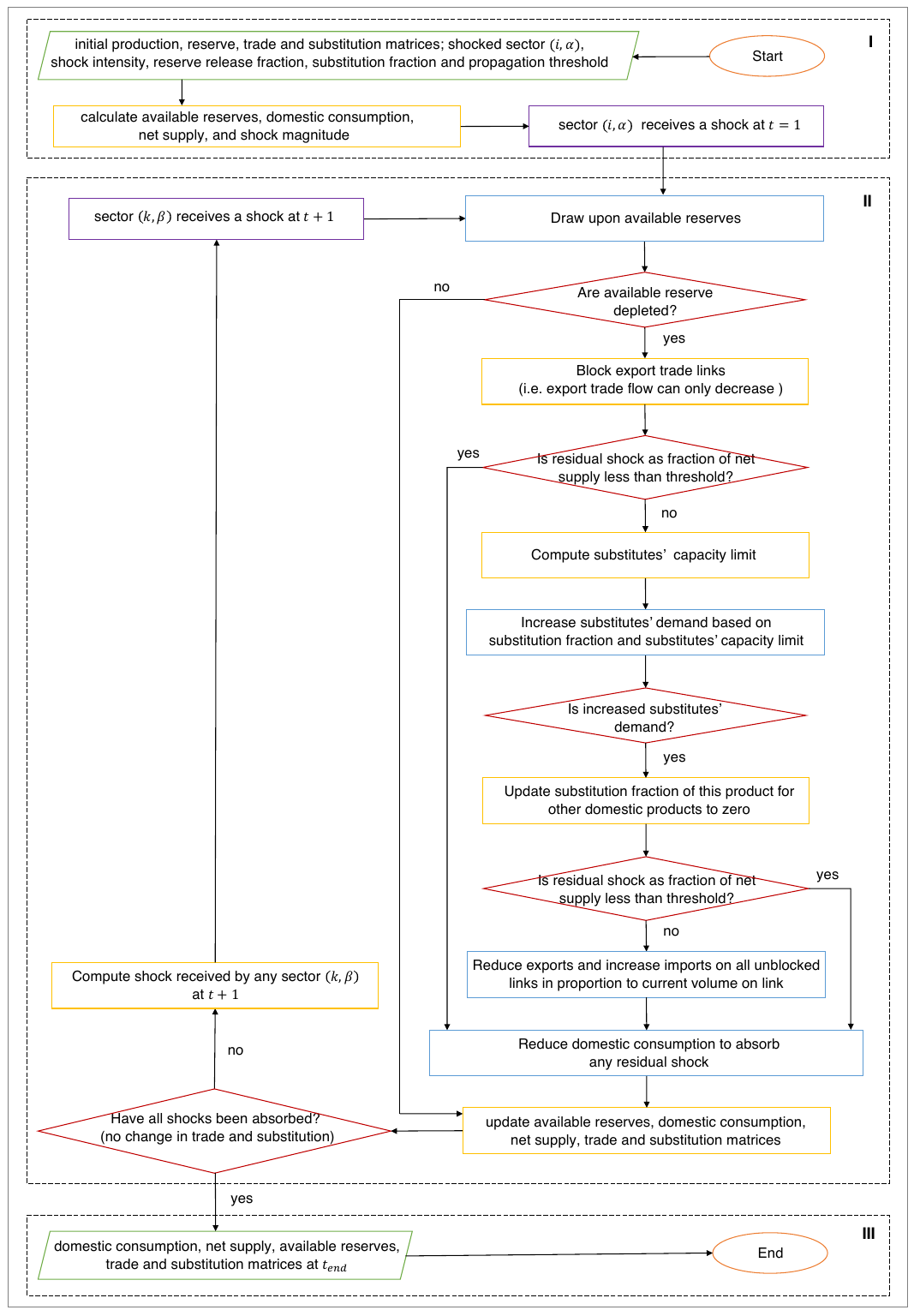}
\caption{Model flow chart. The model comprises three sequential modules: I Shock Generation, II Iterative Propagation, and III Results Output. Sector ($i$, $\alpha$) denotes product $\alpha$ in country $i$. Orange ellipses mark the model's start and end points, green parallelograms denote inputs/outputs, gold rectangles denote calculations/updates, purple rectangles denote that one sector receives a shock at iteration $t$, blue rectangles denote strategy response, red diamonds denote conditional judgments, and black unidirectional arrows indicate flow direction.}
\label{fig5}
\end{figure}

A shock to production induces a short-term out-of-equilibrium condition in which food supply shortfalls are resolved via local adjustments or trade relationships \cite{26}. From a mass balance perspective, supply shortages at the country level may be compensated by drawing on reserves, increasing demand for substitute products, adjusting trade flows, and reducing consumption. The proposed model simulates shock propagation through these response processes using three sequential modules, while preserving mass balance at the country level (Fig. \ref{fig5}).

\bigskip
\textbf{Module I:} Shock Generation. A reduction in the production of a particular product in the target country is set, generating the initial shock. The magnitude of this initial shock is obtained as the product of the country's production for that product and the shock intensity parameter $f_p$ (satisfying $0 < f_p \leq 1$).

\textbf{Module II:} Iterative Propagation. The shock propagates through multiple iterations within the multilayer trade network. In each iteration (denoted as iteration step $t$), the affected country sequentially activates four response strategies for the shortage product:

(1) Reserve release. Available reserves of the product are first drawn upon to directly offset supply shortfalls, absorbing the shock. The available reserve volume is determined by the product of its initial reserve and its reserve release fraction (the release fraction of country $i$ for product $\alpha$ is denoted as $f_{r, i}^{\alpha}$, satisfying $0 \leq f_{r,i}^\alpha \leq 1$).

(2) Cross-product substitution. After available reserves are depleted, demand for substitutes of the product is increased, with the impact spread to each substitute link according to substitution fractions (the fraction of country $i$ using product $\beta$ to substitute for product $\alpha$ is denoted as $f_{s, i}^{\beta \alpha}$, satisfying $0 \leq f_{s, i}^{\beta \alpha} \leq 1$). This propagates the shock to domestic substitutes of the product. The increase in substitution demand is subject to dual constraints of the substitute's capacity limit (the sum of its own available reserves and adjustable trade volume) and competitive substitution demand from multiple products. If total demand for the same substitute product from multiple products in shortage exceeds its capacity limit, the capacity is allocated to each shortage product in proportion to its respective substitution demand. The exception to this strategy is that a product whose demand for substitutes has already been increased following a shock cannot then serve as a substitute for other domestic products.

(3) Trade adjustment. The trade balance for the product is reduced by decreasing exports and increasing imports, with the impact spread to each trade link proportionally to the current trade volume on that link. This propagates the shock to the country's trading partners for the product. The exception to this strategy is when a trading partner's available reserves of the product are depleted, in which case its export links are blocked and imports cannot be increased from that partner.

(4) Consumption reduction. Any shock that could not be propagated is ultimately absorbed by reducing domestic consumption of the product.

A propagation threshold $\rho$ is introduced, allowing the country to directly absorb minor shocks not exceeding $\rho$ fraction of current supply through consumption reduction, bypassing cross-product substitution and/or trade adjustment. The aforementioned four steps are repeated until all shocks are absorbed, concluding the Module II iteration.

\textbf{Module III:}  Results Output. The core variables at iteration termination (denoted as $t_{end}$), such as domestic consumption of each product in each country, are output to measure shock impact. 

For a detailed description of the model and an overview of the parameters and variables, see Section~\ref{model} and Table \ref{tab1}.

\bigskip
After the simulation concludes, we verify that the variable changes for any product $\alpha$ in any country $i$ satisfy: $\Delta P_i^\alpha + \Delta I_i^\alpha = \Delta E_i^\alpha + \Delta R_i^\alpha + \Delta C_i^\alpha +  \Delta O_{E,i}^\alpha - \Delta O_{I,i}^\alpha$. Compared to the initial mass balance, the right-hand side includes two additional terms resulting from the cross-product substitution response: the increased demand for the product itself $\Delta O_{E,i}^\alpha$, and increased substitution demand of the product $\Delta O_{I,i}^\alpha$. At the country level, we verify that the total changes across all products in any country $i$ satisfy: $\Delta P_i + \Delta I_i = \Delta E_i + \Delta R_i + \Delta C_i$. Furthermore, we verified that the sum of the changes in reserves $\Delta R$ and changes in consumption $\Delta C$ for all products across all countries matches the magnitude of the initial shock. These three validation equations hold within a tolerance level of $\rho$.

In our response model, reserve release and consumption reduction allow shocks to be absorbed locally; cross-product substitution induces the cross-layer propagation of shocks to substitute product layers and forms the back-propagating cross-layer feedback; and trade adjustment causes shocks to propagate within each trade layer, ultimately affecting both direct and indirect trading partners of the initially shocked country across all layers. To quantify the effect of cross-product substitution on shock mitigation and propagation, a no-substitution baseline is established by setting any substitution fraction $f_{s, i}^{\beta \alpha}=0$. Under this configuration, cross-layer propagation is prohibited, and shocks can only propagate along trade links within the shocked product layer, impacting the trading countries therein.

In this simulation study, to simplify parameter sensitivity analysis, we assume uniform reserve release fractions ($f_{r, i}^{\alpha}=f_r$) of all countries for all products and uniform substitution fractions ($f_{s,i}^{\beta \alpha} = f_{s,i}^{\alpha \beta} = f_s$) between any two layers across all countries. This yields four input parameters for the model run: shock intensity $f_p$, reserve release fraction $f_r$, substitution fraction $f_s$, and propagation threshold $\rho$. We set $f_r=0.5$ (see Fig. \ref{fig6} for sensitivity analysis) and $\rho=0.001\%$ for all simulations, while configuring $f_p$ and $f_s$ according to specific simulation scenarios. The model typically converges within 20 iterations under this configuration, with rare exceptions only in specific shock scenarios.

\subsection{Quantifying shock impacts and substitution effects}\label{sub:index}

The consumption deficit arising from consumption reduction responses constitutes a core metric for assessing food security threats \cite{26}. The impact severity of a production shock originating from product $\alpha$ in country $i$ on product $\beta$ in country $k$, denoted as $D_{i k}^{\alpha \beta}$, is defined as:

\begin{equation}
D_{ik}^{\alpha\beta} = \Delta C_{ik}^{\alpha\beta} / C_{k,0}^\beta,
\label{eq2}
\end{equation}

\begin{equation}
\Delta C_{ik}^{\alpha\beta} = C_{ik,end}^{\alpha\beta} - C_{k,0}^{\beta}.
\label{eq3}
\end{equation}
Where $C_{k,0}^\beta$ and $C_{ik,end}^{\alpha\beta}$ represent pre-shock and post-shock consumption of product $\beta$ in country $k$, respectively. $D_{i k}^{\alpha \beta}<0$ indicates a consumption deficit, with more negative values corresponding to greater severity.

Building upon this metric, impact indicators can be constructed for individual countries, product layers, and the entire multilayer network. The cumulative deficit on country $k$ across $M$ product types, denoted as $D_{i k}^{\alpha M}$, is:

\begin{equation}
D_{ik}^{\alpha M} = \frac{\sum_\beta^M \Delta C_{ik}^{\alpha\beta}}{\sum_\beta^M C_{k,0}^{\beta}}.
\label{eq4}
\end{equation}
The mean deficit $D_{i N}^{\alpha\beta}$ and deficit unevenness $U_{i N}^{\alpha\beta}$ for product $\beta$ across the network are:

\begin{equation}
D_{iN}^{\alpha\beta} = \mu_k \left(D_{ik}^{\alpha\beta}\right),
\label{eq5}
\end{equation}

\begin{equation}
U_{iN}^{\alpha\beta} = \sigma_k \left(D_{ik}^{\alpha\beta}\right).
\label{eq6}
\end{equation}
Similarly, for the multilayer network:

\begin{equation}
D_{iN}^{\alpha M} = \mu_k \left(D_{ik}^{\alpha M}\right),
\label{eq7}
\end{equation}

\begin{equation}
U_{iN}^{\alpha M} = \sigma_k \left(D_{ik}^{\alpha M}\right).
\label{eq8}
\end{equation}
Here, $\mu_k$ and $\sigma_k$ denote the mean and standard deviation across all $N$ countries, respectively. $U_{iN}^{\alpha\beta} >0$ indicates that the deficit magnitude of product $\beta$ varies across countries, with larger values corresponding to greater impact inequality. Likewise, $U_{iN}^{\alpha M}>0$ indicates cross-country variation in the multilayer network.

\bigskip
By comparing baseline (no substitution/two-layer substitution) and analytical (two-layer/three-layer) simulation results, the effects of cross-product substitution and the increased product layers can be quantified. The differential deficit on product $\beta$ in country $k$ is:

\begin{equation}
\Delta D_{ik}^{\alpha\beta} = D_{ik}^{\alpha\beta} - \bar{D}_{ik}^{\alpha\beta}.
\label{eq9}
\end{equation}
Where $\bar{D}_{ik}^{\alpha\beta}$ and $D_{ik}^{\alpha\beta}$ denote the consumption deficit on product $\beta$ in country $k$ under the baseline and analytical simulations, respectively. $\Delta D_{ik}^{\alpha\beta}>0$ indicates analytical simulations compensated consumption deficit, whereas $\Delta D_{ik}^{\alpha\beta}<0$ indicates amplified deficit. When the baseline deficit is non-zero ($\bar{D}_{ik}^{\alpha\beta} \neq 0$), the deficit compensation rate of the analytical simulation is calculated as:

\begin{equation}
RD_{ik}^{\alpha\beta} = \frac{\Delta D_{ik}^{\alpha\beta}}{|\bar{D}_{ik}^{\alpha\beta}|} \times 100\%.
\label{eq10}
\end{equation}
Here, $RD_{ik}^{\alpha\beta}>0$ indicates effective deficit compensation, with larger values corresponding to greater efficiency; conversely, $RD_{ik}^{\alpha\beta}<0$ indicates deficit amplification, with more negative values representing greater exacerbation.

Similarly, the effect on country $k$ from the analytical simulation relative to the baseline is:

\begin{equation}
\Delta D_{ik}^{\alpha M} = D_{ik}^{\alpha M} - \bar{D}_{ik}^{\alpha M},
\label{eq11}
\end{equation}

\begin{equation}
RD_{ik}^{\alpha M} = \frac{\Delta D_{ik}^{\alpha M}}{|\bar{D}_{ik}^{\alpha M}|} \times 100\%.
\label{eq12}
\end{equation}
The effect on the $\beta$ product layer is:

\begin{equation}
\Delta D_{iN}^{\alpha \beta} = D_{iN}^{\alpha \beta} - \bar{D}_{iN}^{\alpha \beta},
\label{eq13}
\end{equation}

\begin{equation}
\Delta U_{iN}^{\alpha \beta} = \bar{U}_{iN}^{\alpha \beta} - U_{iN}^{\alpha \beta},
\label{eq14}
\end{equation}

\begin{equation}
RD_{iN}^{\alpha\beta} = \frac{\Delta D_{iN}^{\alpha\beta}}{|\bar{D}_{iN}^{\alpha\beta}|} \times 100\%,
\label{eq15}
\end{equation}

\begin{equation}
RU_{iN}^{\alpha\beta} = \frac{\Delta U_{iN}^{\alpha\beta}}{\bar{U}_{iN}^{\alpha\beta}} \times 100\%.
\label{eq16}
\end{equation}
Here, $\Delta U_{iN}^{\alpha\beta}$ represents the differential impact on deficit unevenness; $\Delta U_{iN}^{\alpha\beta}>0$ indicates reduced impact inequality, whereas $\Delta U_{iN}^{\alpha\beta}<0$ indicates aggravated inequality. When $\bar{U}_{iN}^{\alpha\beta} \neq 0$, the deficit unevenness reduction rate $RU_{iN}^{\alpha\beta}$ of the analytical simulation is calculated. Specifically, $RU_{iN}^{\alpha\beta}>0$ indicates effective inequality reduction, with larger values corresponding to higher efficiency; $RU_{iN}^{\alpha\beta}<0$ indicates inequality aggravation, with more negative values corresponding to greater aggravation. Similarly, the effect on the multilayer network is:

\begin{equation}
\Delta D_{iN}^{\alpha M} = D_{iN}^{\alpha M} - \bar{D}_{iN}^{\alpha M},
\label{eq17}
\end{equation}

\begin{equation}
\Delta U_{iN}^{\alpha M} = \bar{U}_{iN}^{\alpha M} - U_{iN}^{\alpha M},
\label{eq18}
\end{equation}

\begin{equation}
RD_{iN}^{\alpha M} = \frac{\Delta D_{iN}^{\alpha M}}{|\bar{D}_{iN}^{\alpha M}|} \times 100\%,
\label{eq19}
\end{equation}

\begin{equation}
RU_{iN}^{\alpha M} = \frac{\Delta U_{iN}^{\alpha M}}{\bar{U}_{iN}^{\alpha M}} \times 100\%.
\label{eq20}
\end{equation}

\bigskip
When assessing the effects of cross-product substitution and the increased product layers, a single model simulation in this study is configured as an exhaustive shock protocol, wherein production shocks are applied separately to each country in the shocked $\alpha$ product layer. Under this configuration, the cumulative impact severity for product $\beta$ in country $k$ is defined as:

\begin{equation}
D_{Nk}^{\alpha\beta} = \frac{\sum_i^N \Delta C_{ik}^{\alpha\beta}}{C_{k,0}^{\beta}}.
\label{eq21}
\end{equation}
Based on this metric, the computational logic of equations \ref{eq4}--\ref{eq20} can be adopted to calculate the impact of a single simulation for countries, product layers, and the entire multilayer network ($D_{Nk}^{\alpha M}$, $D_{NN}^{\alpha\beta}$, $U_{NN}^{\alpha\beta}$, $D_{NN}^{\alpha M}$, $U_{NN}^{\alpha M}$), as well as its effects relative to the baseline simulation ($\Delta D_{Nk}^{\alpha\beta}$, $RD_{Nk}^{\alpha\beta}$, $\Delta D_{Nk}^{\alpha M}$, $RD_{Nk}^{\alpha M}$, $\Delta D_{NN}^{\alpha\beta}$, $\Delta U_{NN}^{\alpha\beta}$, $RD_{NN}^{\alpha\beta}$, $RU_{NN}^{\alpha\beta}$, $\Delta D_{NN}^{\alpha M}$, $\Delta U_{NN}^{\alpha M}$, $RD_{NN}^{\alpha M}$, $RU_{NN}^{\alpha M}$).

\subsection{Data}\label{sub:data}

As a major component of global food trade, cereals represent an ideal entry point for investigating systemic food security risks. Production, ending reserves, and trade volumes for each primary product and its secondary products within the cereal group are converted to kcal equivalents and aggregated by primary product category, country, and year, forming an annual country-scale dataset encompassing eight major cereal products (wheat, rice, barley, maize, rye, oats, millet, sorghum). Product list and conversion coefficients are detailed in Table \ref{tab3}.

To analyze the effects of cross-product substitution and the increased product layers, the model is initialized using 2017 cereal data to exclude interference from shock events such as China-US trade friction, the COVID-19 pandemic, and the Russia-Ukraine conflict, thereby ensuring the stability of the trade network. To evaluate the cross-product substitution effectiveness under realistic shock scenarios, cereal data from the year preceding three shock events are selected: 2022 rice data for the Indian rice export ban scenario, 2021 wheat data for the Ukrainian wheat export disruption scenario, and 2011 maize data for the U.S. maize production reduction scenario.

Production and trade data are sourced from the Food and Agriculture Organization of the United Nations' online database (FAOSTAT, fao.org/faostat/), and reserve data from the Production, Supply and Distribution database of the United States Department of Agriculture Foreign Agricultural Service (USDA-PSD, apps.fas.usda.gov/psdonline/). We acquired data spanning 31 years from 1993 to 2023 in August 2025. There are only small changes to the set of countries after 1992, facilitating analysis, whereas production and trade data were only updated through 2023 at the time of data collection. For trade data, if discrepancies exist in the trade amounts between two countries, the average value was used; if only one country reported that trade occurred, the single reported value was adopted \cite{41}. Country and product names in the reserve data were rectified to match the FAOSTAT data. Since USDA-PSD reports aggregate reserves for the EU-15 (pre-1999), EU-27 plus United Kingdom (1999--2015), and EU-27 (2016 and after), these reserves were allocated to individual EU countries annually based on their respective shares of cereal production \cite{26}. FAOSTAT population data were used to subset the trade network, considering only countries with populations exceeding half a million people during the period 1993--2023, yielding data for 171 countries \cite{26}. The country list is provided in Tables \ref{tab4}--\ref{tab8}.

\backmatter

\bmhead{Acknowledgements}

This work was supported by the National Natural Science Foundation of China (Grant No. 72371031).

\section*{Declarations}

\begin{itemize}
\item Conflict of interest/Competing interests

The authors declare no competing interests. 
\item Data availability 

The data used in this study as input for simulations are available on GitHub at https://github.com/guo2877/multiNet-shock-response.
\item Code availability 

Simulation and analysis code for this study is available on GitHub at https://github.com/guo2877/multiNet-shock-response.
\item Author contributions

Y.F. and F.G. contributed to study conception and design. F.G. collected the data. F.G. undertook the formal analyses. Y.F., J.Z. and F.G. interpreted the results. F.G. and J.Z. visualized the data.  F.G. wrote the first draft.  Y.F., L.Q. and J.Z. contributed to the review and editing of the paper.
\end{itemize}

\bibliography{sn-bibliography}


\newpage
\section{Supplementary information}

\subsection{Shock response model details}\label{model}
In our model, we define the annual net supply $S_i^\alpha$ of a food product $\alpha$ in country $i$ as:

\begin{equation}
S_i^\alpha = P_i^\alpha + I_i^\alpha - E_i^\alpha.
\label{eq22}
\end{equation}
Where, $P_i^\alpha$ is the production of product $\alpha$ in country $i$, $I_i^\alpha$ is the import volume, and $E_i^\alpha$ is the export volume. The difference between this net supply and domestic consumption $C_i^\alpha$ result in a change in the reserve $\Delta R_i^\alpha$, so that:
\begin{equation}
S_i^\alpha = C_i^\alpha + \Delta R_i^\alpha.
\label{eq23}
\end{equation}
Here, a positive value of $\Delta R_i^\alpha$ indicates a transfer to reserves, i.e., an increase in reserves. We assumed that initial consumption equals net supply, implying that $\Delta R_i^\alpha=0$ before any shock.

The model is initialized with realistic production, reserves, and trade data. The trade data is represented by a matrix $A^\alpha$, where element $a_{ik}^\alpha$ is the export volume of product $\alpha$ from country $i$ to country $k$. Cross-product substitution data is represented by a matrix $O^{\alpha \beta}$, wherein element $o_i^{\beta \alpha}$ is the substitution volume of product $\alpha$ by $\beta$ in country $i$. We assumed that initially any $o_i^{\beta \alpha}=0$, implying that no intra-national product substitution occurs before any shock. The trade and substitution matrices for the $M$ types product constitute the supra-adjacency matrix $G$.

\bigskip
In the first iteration of the model ($t$=1), a supply shock is initiated from a drop in the production of one product:
\begin{equation}
\Delta \tilde{S}_{i,1}^\alpha = \Delta P_i^\alpha = -f_p P_i^\alpha.
\label{eq24}
\end{equation}
Where $\Delta P_i^\alpha = 0$ for all countries but one or a few affected countries by the initial shock, and the shock intensity parameter $f_p$ (satisfying $0 < f_p \leq 1$) can be used to control the magnitude of the initial shock. In the following iterations, $\Delta \tilde{S}_i^\alpha$ is caused by a decrease in imports or an increase in exports within the product layer, as well as an increase in substitution demands across layers, as the shock is propagated within layers through trade and across layers through cross-product substitution. The notation ($\sim$) is used to distinguish different temporary values of the shock within a single iteration.

\bigskip
At the start of an iteration, all countries receiving a new shock to any product $\alpha$ (from any source) first attempt to absorb it through their available reserves of product $\alpha$:
\begin{equation}
\Delta R_{i,t}^\alpha = \max\left(\Delta \tilde{S}_{i,t}^\alpha, -R_{i,t}^\alpha\right).
\label{eq25}
\end{equation}
Where the initial available reserve $R_{i,0}^\alpha$ is derived as the product of its reserve release fraction $f_{r,i}^\alpha$ (satisfying $0 \leq f_{r,i}^\alpha \leq 1$) and the ending reserve data. Any country that has depleted its available reserves of product $\alpha$ (or initially has none) will block all its export trade flows in the product $\alpha$ trade layer, meaning it cannot increase exports of product $\alpha$. This country is removed from the set $\mathcal{U}^\alpha$, as shown in equation \ref{eq26}, which initially contains all countries in the product $\alpha$ trade layer.
\begin{equation}
\mathcal{U}^\alpha = \mathcal{U}^\alpha \setminus \{i \mid \Delta \tilde{S}_{i,t}^\alpha < -R_{i,t}^\alpha\}.
\label{eq26}
\end{equation}
At this point, the residual shock $\Delta \tilde{\tilde{S}}_{i,t}^\alpha$ can be propagated to domestic substitutes of product $\alpha$ through cross-product substitution:
\begin{equation}
\Delta \tilde{\tilde{S}}_{i,t}^\alpha = \Delta \tilde{S}_{i,t}^\alpha - \Delta R_{i,t}^\alpha.
\label{eq27}
\end{equation}
We define the propagation threshold $\rho$ so that countries directly absorb the residual shock by reducing domestic consumption when $|\Delta \tilde{\tilde{S}}_{i,t}^\alpha| < \rho S_{i,t}^\alpha$ instead of increasing substitution demand.

Considering the physically available volume of a product in the short-term, we set the maximum volume of the product  $\beta$ currently available for substitution, i.e., its substitution capacity limit, as:
\begin{equation}
L_{i,t}^\beta = \left(R_{i,t}^\beta + \Delta R_{i,t}^\beta\right) + \left(\sum_k a_{ik,t}^\beta + \sum_k a_{ki,t}^\beta \cdot \mathbf{1}_{\mathcal{U}^\beta}(k)\right).
\label{eq28}
\end{equation}
Where the first parenthesis calculates the current available reserves, and the second parenthesis calculates the total volume on all the country's trade links within the product $\beta$ trade layer, excluding any ``blocked'' import links. If country $k$ is not in $\mathcal{U}^\beta$ , then $\mathbf{1}_{\mathcal{U}^\beta}(k)$  takes the value of 1, and 0 otherwise. The total increased demand for a substitute product from multiple shocked products cannot exceed its substitution capacity limit.  Therefore, the actual increased demand for any substitute product $\beta$, $\Delta O_{E,i,t}^\beta$, is given by:
\begin{equation}
\Delta O_{E,i,t}^\beta = \min\left(L_{i,t}^\beta, -\sum_\alpha f_{s,i}^{\beta\alpha} \Delta \tilde{\tilde{S}}_{i,t}^\alpha\right).
\label{eq29}
\end{equation}
Where $f_{s,i}^{\beta\alpha} \Delta \tilde{\tilde{S}}_{i,t}^\alpha$ is the substitution demand of shocked product $\alpha$ for $\beta$, and $f_{s,i}^{\beta\alpha}$ denotes the substitution fraction of product $\beta$ for product $\alpha$ (satisfying $0 \leq \sum_\beta f_{s,i}^{\beta\alpha} \leq 1$). Each country then changes the substitution volume (+ for increasing substitution) on each substitution link proportionally to the substitution demand of the shocked product. From this, the change in any substitution link can be calculated as:
\begin{equation}
\Delta o_{i,t}^{\beta\alpha} = \frac{f_{s,i}^{\beta\alpha} \Delta \tilde{\tilde{S}}_{i,t}^\alpha}{\sum_\alpha f_{s,i}^{\beta\alpha} \Delta \tilde{\tilde{S}}_{i,t}^\alpha} \Delta O_{E,i,t}^\beta.
\label{eq30}
\end{equation}
The increased substitution demand of shocked product $\alpha$ is:
\begin{equation}
\Delta O_{I,i,t}^\alpha = \sum_\beta \Delta o_{i,t}^{\beta\alpha}.
\label{eq31}
\end{equation}
When $\Delta O_{I,i,t}^\alpha > 0$, product $\alpha$ cannot substitute for other domestic products (preventing mutual substitution effects that would cause shocks to propagate interactively among a country's products and hinder model convergence), and the substitution fraction $f_{s,i}^{\alpha\beta}$ of product $\alpha$ for other domestic products is updated to 0.

If, after cross-product substitution propagation, the residual shock still exceeds $\rho S_{i,t}^\alpha$, it can be propagated through trade to other countries within the product $\alpha$ trade layer. A country can transfer a shock to its trading partners for product $\alpha$ by reducing exports or increasing imports, that is, by decreasing its balance of trade $T^\alpha$. The upper bound on the magnitude of this change $\Delta T_{i,t}^\alpha$ is capped at the total volume $V^\alpha$ on all the country's trade links in the product $\alpha$ trade layer:
\begin{equation}
V_{i,t}^\alpha = \sum_k a_{ik,t}^\alpha + \sum_k a_{ki,t}^\alpha \cdot \mathbf{1}_{\mathcal{U}^\alpha}(k),
\label{eq32}
\end{equation}
\begin{equation}
\Delta T_{i,t}^\alpha = \max\left(\Delta \tilde{\tilde{S}}_{i,t}^\alpha + \Delta O_{I,i,t}^\alpha, -V_{i,t}^\alpha\right).
\label{eq33}
\end{equation}
Each country then changes the trade volume on each unblocked link (- for exports, + for imports) proportionally to the current volume on that link. From this, the change in any trade link can be calculated as:
\begin{equation}
\Delta a_{ik,t}^\alpha = \left[\left(\frac{\Delta T_{i,t}^\alpha}{V_{i,t}^\alpha}\right) - \mathbf{1}_{\mathcal{U}^\alpha}(i)\left(\frac{\Delta T_{k,t}^\alpha}{V_{k,t}^\alpha}\right)\right] a_{ik,t}^\alpha.
\label{eq34}
\end{equation}
Finally, the portion of the initial shock not absorbed by reserves, or propagated through inter-layer substitution or intra-layer trade results in a decrease in consumption:
\begin{equation}
\Delta C_{i,t}^\alpha = \Delta \tilde{S}_{i,t}^\alpha - \Delta R_{i,t}^\alpha + \Delta O_{I,i,t}^\alpha - \Delta T_{i,t}^\alpha.
\label{eq35}
\end{equation}

\bigskip
To start the next iteration, we update the reserves $R$, consumption $C$, net supply $S$, trade $A$ and substitution $O$ matrix based on the computed $\Delta$. For a given product in a country, the actual change in $S^\alpha$ only depends on the portion of the shock that was absorbed internally:
\begin{equation}
\Delta S_{i,t}^\alpha = \Delta R_{i,t}^\alpha + \Delta C_{i,t}^\alpha.
\label{eq36}
\end{equation}

The new shock received by each country in each product trade layer at the next iteration is determined by three changes occurring in the current iteration: (i) the decrease in imports induced by other countries' actions within the same layer; (ii) the increase in exports triggered by other countries' actions within the same layer; and (iii) the demand increase arising from inter-layer substitution from other products within that country. It can be calculated as:
\begin{equation}
\Delta \tilde{S}_{i,t+1}^\alpha = \sum_k \Delta a_{ki,t}^\alpha - \sum_k \Delta a_{ik,t}^\alpha - \sum_\beta \Delta o_{i,t}^{\alpha\beta} + \Delta T_{i,t}^\alpha.
\label{eq37}
\end{equation}
Where the last term $\Delta T_{i,t}^\alpha$ serves to remove the impact of the country's own trade actions. When $\Delta \tilde{S}_{i,t+1}^\alpha = 0$ for all countries in all trade layers, all shocks have been internally absorbed and the simulation ends.

The model does not preset the number of iterations or make specific assumptions about the duration of a single iteration, but assumes that shock propagation stops in a relatively short period of time, shorter than the data time scale (annual), during which production cannot increase \cite{25,27,28,30}.

\newpage
\subsection{Parameters and variables}\label{param}
The following Table \ref{tab1} lists all parameters and variables in the model and indicates their sources and states. Sources include those directly derived from data or obtained through iterative updating calculations. States include static or dynamically changing over time. 

\newpage
\begin{table}[ht]
\centering
\caption{All symbols used in the model, ordered origin.}\label{tab1}%
\begin{tabularx}{\columnwidth}{@{}cXcc@{}}  
\toprule
Symbol & Description & Origin & State\\
\midrule
$M$ & Number of product types in multilayer trade network & data & static \\
$N$ & Number of countries in multilayer trade network & data & static \\
$P_i^\alpha$ & Production of product $\alpha$ in country $i$ & variable & dynamic \\
$R_i^\alpha$ & Available reserves of product $\alpha$ in country $i$ & variable & dynamic \\
$A^\alpha$ & Trade matrix of product $\alpha$ $(a_{ik}^\alpha = \text{export volume of product } \alpha \text{ from country } i \text{ to } k)$ & variable & dynamic \\
$O^\alpha$ & Substitution matrix between products $\alpha$ and $\beta$ $(o_i^{\beta\alpha} = \text{substitution volume of product } \alpha \text{ by } \beta \text{ in country } i)$ & variable & dynamic \\
$G$ & Super-adjacency matrix for trade and substitution $(\text{composed of } A^\alpha \text{ and } O^\alpha \text{ for } M \text{ types product})$ & variable & dynamic \\
$E_i^\alpha$ & Exports of product $\alpha$ in country $i$ , i.e. $\sum_ka_{ik}^\alpha$ & variable & dynamic \\
$I_i^\alpha$ & Imports of product $\alpha$ in country $i$, i.e. $\sum_ka_{ki}^\alpha$ & variable & dynamic \\
$O_{E,i}^\alpha$ & Substitution volume of product $\alpha$ in country $i$ to substitute for other products, i.e. $\sum_\beta o_i^{\alpha\beta}$ & variable & dynamic \\
$O_{I,i}^\alpha$ & Substitution volume of other products in country $i$ to substitute for $\alpha$, i.e. $\sum_\beta o_i^{\beta\alpha}$ & variable & dynamic \\
$S_i^\alpha $ & Net supply of product $\alpha$ in country $i$, $S_i^\alpha = P_i^\alpha + I_i^\alpha - E_i^\alpha$ & variable & dynamic \\
$C_i^\alpha $ & Consumption of product $\alpha$ in country $i$ & variable & dynamic \\
$f_p$ & Magnitude of initial shock as a fraction of affected country's $P_i^\alpha$ & parameter & static \\
$f_{r,i}^\alpha$ & Fraction of country $i$'s actual reserves of product $\alpha$ that are available to absorb shocks & parameter & static \\
$f_{s,i}^{\alpha\beta}$ & Fraction of country $i$ using product $\beta$ to substitute for $\alpha$ & parameter & static \\
$\rho$ & Minimum threshold for shock propagation (as a fraction of $S_i^\alpha $) & parameter & static \\
$t$ & A specific iteration & calculated & dynamic \\
$\Delta \tilde{S}_{i,t}^\alpha$ & Shock magnitude received by product $\alpha$ in country $i$ & calculated & dynamic \\
$\mathcal{U}^\alpha$ & Countries set where the exports of product $\alpha$ cannot be increased & calculated & dynamic \\
$L_{i,t}^\alpha$ & Substitution capacity limit of product $\alpha$ in country $i$ & calculated & dynamic \\
$V_{i,t}^\alpha$ & Upper bound of trade change of product $\alpha$ in country $i$ & calculated & dynamic \\
$\Delta R_{i,t}^\alpha$ & {Reserves change of product $\alpha$ in country $i$ (positive value indicates an increase in reserves)} & calculated & dynamic \\
$\Delta O_{I,i,t}^\alpha$ & Increased substitution demand of product $\alpha$ in country $i$ & calculated & dynamic \\
$\Delta O_{E,i,t}^\alpha$ & Increased demand for product $\alpha$ in country $i$ & calculated & dynamic \\
$\Delta T_{i,t}^\alpha$ & Trade change of product $\alpha$ in country $i$ & calculated & dynamic \\
$\Delta C_{i,t}^\alpha$ & Consumption change of product $\alpha$ in country $i$ & calculated & dynamic \\
$\Delta S_{i,t}^\alpha$ & Net supply change of product $\alpha$ in country $i$ & calculated & dynamic \\
\botrule
\end{tabularx}
\end{table}

\newpage
\subsection{Network properties}\label{prop}
Here, we consider the weakly connected component of the trade network for any product $\alpha$ (the largest subnetwork where any node can reach others when link directions are ignored), which contains $N_c^\alpha$ nodes and $N_{tl}^\alpha$ trade links. Calculate the total production $P^\alpha$, total reserves $R^\alpha$, and total trade volume $E^\alpha$ of this component. $R^\alpha$ and $E^\alpha$ jointly measure supply scale, reflecting both the network's buffer capacity for absorbing shocks and its potential upper bound for providing substitute supply to other product trade networks.

The reserve distribution evenness $J^\alpha$ and network density $D^\alpha$ are defined as:
\begin{equation}
p_i^\alpha = \frac{\frac{R_i^\alpha}{S_i^\alpha}}{\sum_i^{N_c^\alpha} \frac{R_i^\alpha}{S_i^\alpha}},
\label{eq38}
\end{equation}

\begin{equation}
J^\alpha = -\frac{1}{\log N_c^\alpha} \sum_i^{N_c^\alpha} p_i^\alpha \log p_i^\alpha \in [0,1],
\label{eq39}
\end{equation}

\begin{equation}
D^\alpha = \frac{N_{tl}^\alpha}{N_c^\alpha \cdot (N_c^\alpha - 1)} \in [0,1].
\label{eq40}
\end{equation}
Where $\frac{R_i^\alpha}{S_i^\alpha}$ is the fraction of reserves $R_i^\alpha$ of country $i$ to its net supply $S_i^\alpha$. $J^\alpha$ approaching 1 indicates more evenness reserve distribution; $D^\alpha$ approaching 1 indicates higher trade path diversification. Whether as a shocked layer or substitute layer, high $J^\alpha$ and $D^\alpha$ ensure coordinated reserve release and switchable trade paths, thereby enabling efficient shock absorption and diversion.

Finally, we calculate the export market concentration (Herfindahl index) $H^\alpha$:
\begin{equation}
H^\alpha = \sum_i^{N_c^\alpha} \left(\frac{\sum_j^{N_c^\alpha} a_{ij}^\alpha}{\sum_i^{N_c^\alpha} \sum_j^{N_c^\alpha} a_{ij}^\alpha}\right)^2 \in [0,1],
\label{eq41}
\end{equation}
where $a_{ij}^\alpha$ is the export volume from country $i$ to country $j$. Higher $H^\alpha$ indicates a stronger export market monopoly, where disturbances in major exporting countries can easily trigger global supply crises.

Supply scale determines supply abundance, reserve distribution evenness and trade path diversification affects supply efficiency, and export market concentration constrains supply stability. These three dimensions jointly influence network supply capacity. Relative deficiency in any dimension may constitute a supply bottleneck, constraining resilience within the network layer and cross-layer substitution capability.

The following Table \ref{tab2} summarizes the above properties for each trade network layer in 2017 as studied in the main text. The wheat network demonstrates relative advantages across all three dimensions. The rice network faces dual constraints of insufficient supply efficiency and high market concentration. The maize network, despite abundant scale, suffers from reduced effective supply capacity due to low trade path diversification combined with high export concentration. The barley network, though smallest in supply scale, benefits from relatively higher supply stability owing to low market concentration.

\newpage
\begin{table}[ht]
\centering
\caption{Basic properties of the trade network for different cereal products. The table reports the total number of network nodes $N$, the number of nodes $N_c^\alpha$ in the weakly connected component, the number of trade links $N_{tl}^\alpha$, total production $P^\alpha$ (kcal), total reserves $R^\alpha$ (kcal), total trade volume $E^\alpha$ (kcal), reserve distribution evenness $J^\alpha$, network density $D^\alpha$, and market concentration $H^\alpha$.}\label{tab2}%
\begin{tabular}{@{}lccccccccc@{}}
\toprule
Net & $N$ & $N_c^\alpha$ & $N_{tl}^\alpha$ & $P^\alpha$ & $R^\alpha$ & $E^\alpha$ & $J^\alpha$ & $D^\alpha$ & $H^\alpha$ \\
 & & & & (kcal) & (kcal) & (kcal) & & & \\
\midrule
Wheat & 171 & 168 & 6974 & $2.58{\times}10^{15}$ & $9.58{\times}10^{14}$ & $7.74{\times}10^{14}$ & 0.89 & 0.25 & 0.07 \\
Rice & 171 & 163 & 2875 & $2.09{\times}10^{15}$ & $5.87{\times}10^{14}$ & $3.14{\times}10^{14}$ & 0.67 & 0.11 & 0.16 \\
Maize & 171 & 162 & 2535 & $4.04{\times}10^{15}$ & $1.21{\times}10^{15}$ & $5.46{\times}10^{14}$ & 0.88 & 0.10 & 0.18 \\
Barley & 171 & 164 & 2139 & $4.93{\times}10^{14}$ & $7.20{\times}10^{13}$ & $1.56{\times}10^{14}$ & 0.79 & 0.08 & 0.09 \\
\botrule
\end{tabular}
\end{table}

\newpage
\subsection{Sensitivity analysis}\label{sens}
\begin{figure}[h]
\centering
\includegraphics[width=0.9\textwidth]{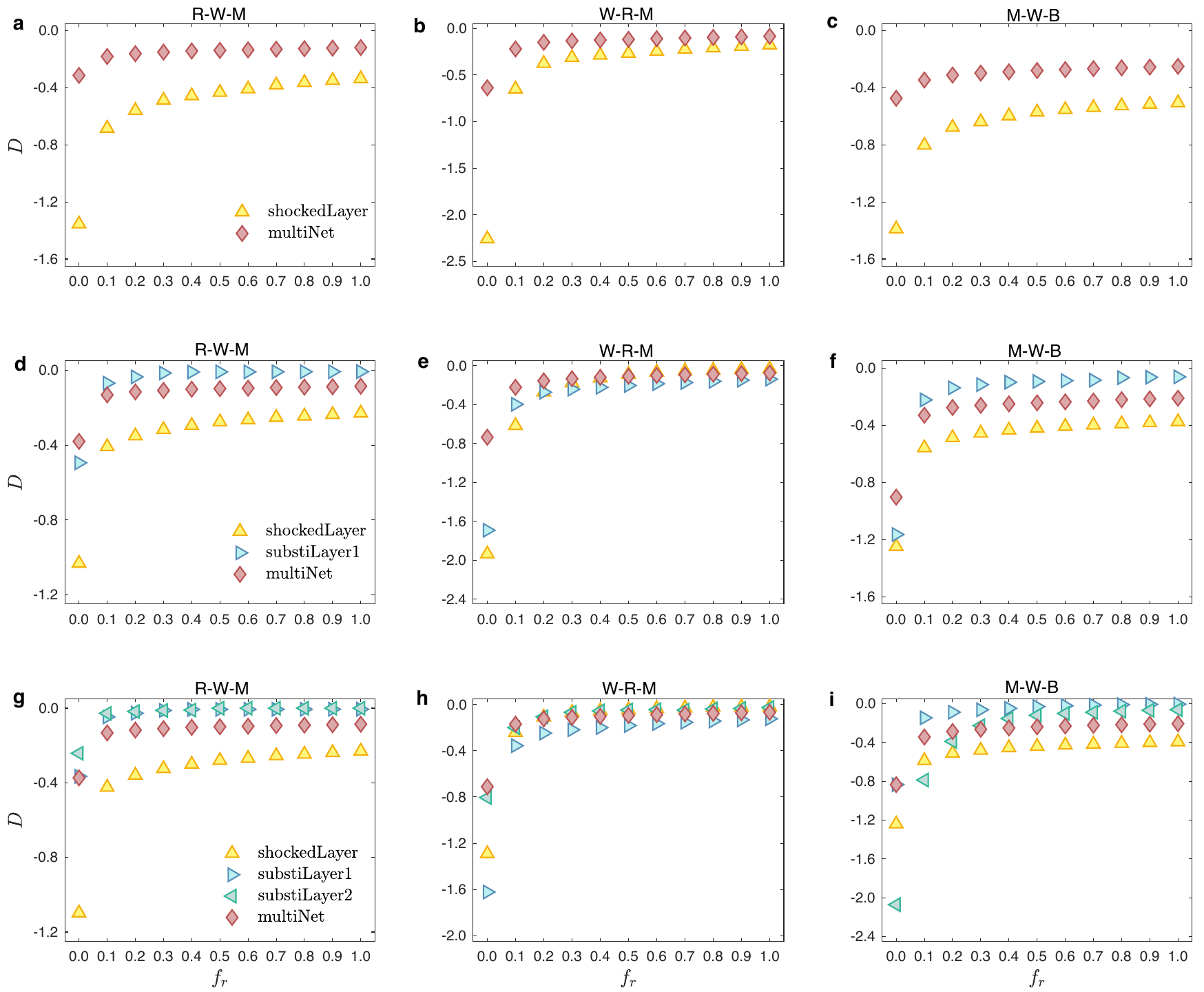}
\caption{Reserve release fraction sensitivity. Fix $f_p=1$ and $f_s=0.2$ (total substitution fraction for shocked layer), impacts of no-substitution ($f_{s1}=f_{s2}=f_{s3}=0$) \textbf{(a-c)}, two-layer substitution ($f_{s1}=0.2$, $f_{s2}=f_{s3}=0$) \textbf{(d-f)}, and three-layer substitution (($f_s1=0.2$, $f_s2=0.1$), ($f_{s1}=f_{s2}=0.1$), ($f_{s1}=f_{s2}=0.1$, $f_{s3}=0.1$)) \textbf{(g-i)} simulations on shocked layer, substitute layer, and the entire three-layer network under $f_r \in [0.0, 1.0]$ (with step size 0.1). No-substitution does not affect substitute layers, thus only displaying average deficits for shocked layer and overall network ($D_{NN}^{\alpha\alpha}$, $D_{NN}^{\alpha M}$). Two-layer substitution does not affect the second substitute layer, thus only displaying average deficits for shocked layer, first substitute layer, and overall network ($D_{NN}^{\alpha\alpha}$, $D_{NN}^{\alpha\beta}$, $D_{NN}^{\alpha M}$). Three-layer substitution displays all ($D_{NN}^{\alpha\alpha}$, $D_{NN}^{\alpha\beta}$, $D_{NN}^{\alpha\gamma}$, $D_{NN}^{\alpha M}$).}
\label{fig6}
\end{figure}

Figure \ref{fig6} presents sensitivity analysis results for the reserve release fraction $f_r$ across the three three-layer network scenarios in Fig. \ref{fig2}a--c. Under no-substitution, two-layer substitution, and three-layer substitution simulations, shock impacts on the shocked layer, substitute layers, and the overall three-layer network decrease monotonically with increasing $f_r$. As $f_r$ increases, countries can draw on more reserves to absorb shocks, thereby reducing impacts on consumption. Beyond $f_r=0.5$, the magnitude of the change in shock impact becomes relatively small. Therefore, we fix $f_r=0.5$ for all impact analyses in the main text simulations.

\newpage

\newpage
\subsection{Effects of cross-product substitution on three-layer network supply risk}\label{effects}
\begin{figure}[h]
\centering
\includegraphics[width=0.9\textwidth]{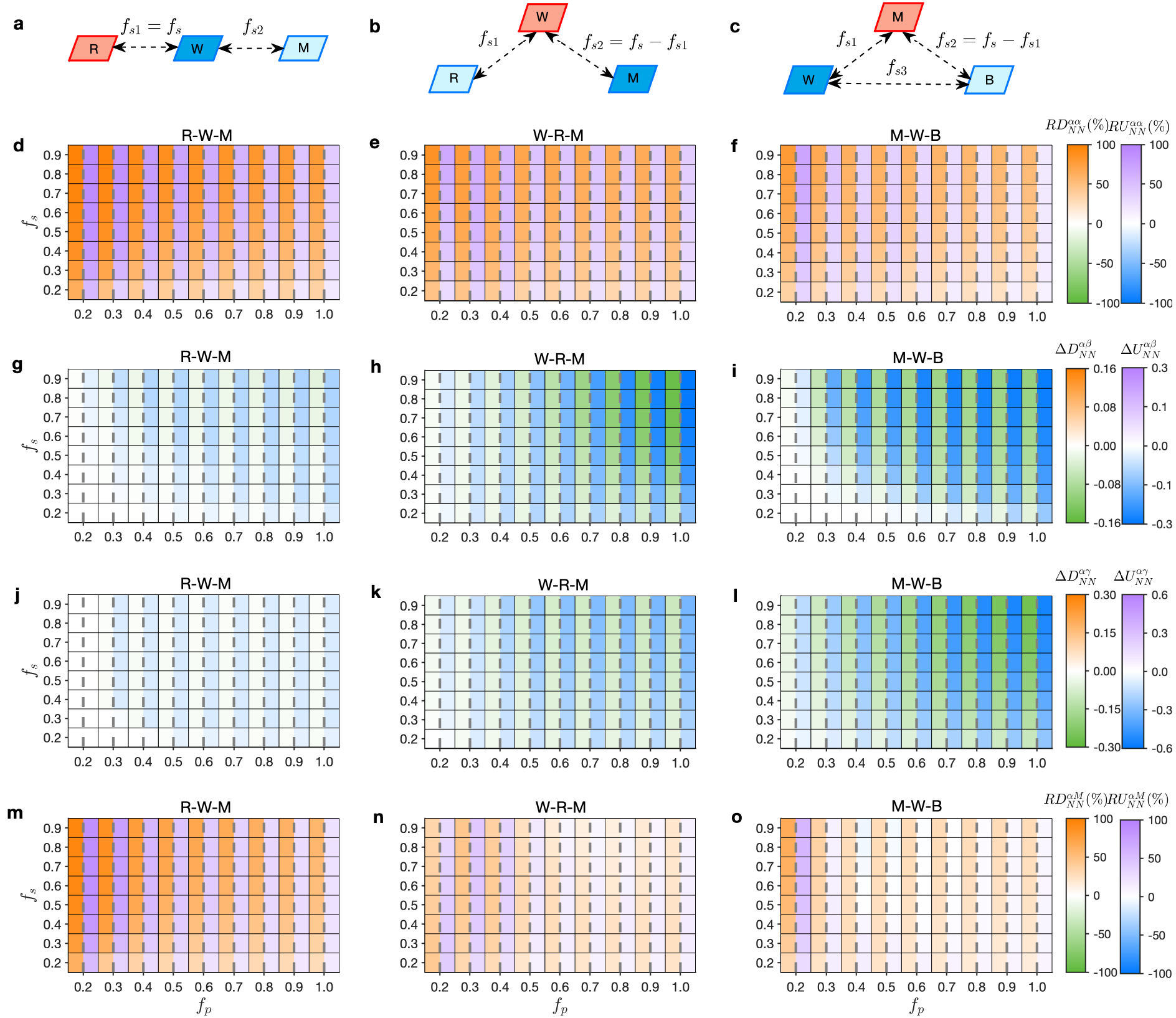}
\caption{Effects of three-layer substitution relative to the no-substitution. \textbf{a-c}, Schematic topology of simulation scenarios. Red quadrilaterals denote the shocked layer, blue quadrilaterals denote the substitute layer (dark blue = strong supply capacity, light blue = weak). Bidirectional dashed lines indicate substitution relationships, where $f_s$ represents the total substitution fraction for shocked layer distributed across $f_{s1}$, $f_{s2}$, and $f_{s3}$. Product layers are abbreviated as R = rice, W = wheat, M = maize, and B = barley, constituting three scenarios: R-W-M chain structure \textbf{(a)}, W-R-M triadic structure \textbf{(b)}, and M-W-B triangular structure \textbf{(c)}. \textbf{d-o}, Under $f_p \in [0.2, 1.0]$ and $f_s \in [0.2, 0.9]$ (both with step size 0.1), effects of three-layer substitution (non-zero ($f_{s1}$, $f_{s2}$) or ($f_{s1}$, $f_{s2}$, $f_{s3}$) combinations) relative to no-substitution ( $f_{s1}=f_{s2}=f_{s3}=0$) on shocked layer \textbf{(d-f)}, first substitute layer (the substitution fraction with the shocked layer is $f_{s1}$) \textbf{(g-i)}, second substitute layer \textbf{(j-l)}, and entire three-layer network \textbf{(m-o)}. Each cell represents the mean effects of all non-zero ($f_{s1}$, $f_{s2}$) or ($f_{s1}$, $f_{s2}$, $f_{s3}$) combinations (step size 0.1) at fixed ($f_p$, $f_s$) and is split into two halves by a gray dashed line. In \textbf{(d-f)} and \textbf{(m-o)}, the left half shows deficit compensation rate ($RD_{NN}^{\alpha\alpha}$, $RD_{NN}^{\alpha M}$), while the right half shows deficit unevenness reduction rate ($RU_{NN}^{\alpha\alpha}$, $RU_{NN}^{\alpha M}$). In \textbf{(g-i)} and \textbf{(j-l)}, the left half indicates differential deficit ($\Delta D_{NN}^{\alpha\beta}$, $\Delta D_{NN}^{\alpha\gamma}$), and the right half shows differential deficit unevenness ($\Delta U_{NN}^{\alpha\beta}$, $\Delta U_{NN}^{\alpha\gamma}$).}
\label{fig7}
\end{figure}

Figure \ref{fig7}a--c illustrates three three-layer network simulation scenarios: (1) rice shocked with wheat and maize substitutes; (2) wheat shocked with rice and maize substitutes; and (3) maize shocked with wheat and barley substitutes. Fig. \ref{fig7}d--l presents the effects of three-layer substitution relative to no-substitution baseline across three scenarios. The effects on shocked and substitute layers exhibit cross-scenario consistency: while cross-product substitution mitigates shock impacts ($RD_{NN}^{\alpha\alpha} > 0$, $RU_{NN}^{\alpha\alpha} > 0$) in the shocked layer, it induces derived impacts ($\Delta D_{NN}^{\alpha\beta} < 0$, $\Delta D_{NN}^{\alpha\gamma} < 0$, and $\Delta U_{NN}^{\alpha\beta} < 0$, $\Delta U_{NN}^{\alpha\gamma} < 0$) in substitute layers. The overall three-layer network displays two main response regimes:

(i) Efficient compensation with inequality reduction ($RD_{NN}^{\alpha M} > 0$, $RU_{NN}^{\alpha M} > 0$): Both average deficit and deficit unevenness declined, representing efficient risk mitigation.

(ii) Efficient compensation with inequality aggravation ($RD_{NN}^{\alpha M} > 0$, $RU_{NN}^{\alpha M} < 0$): Deficits are mitigated on average yet unevenly distributed, concentrating vulnerability in certain countries.

Regime (i) exists in all three scenarios. Regimes (ii) exists only in M-W-B scenario, which may be attributed to the maize layer's limited trade path diversification and high export concentration.

The above effects exhibit consistency with two-layer network conclusions in the main text, corroborating the robustness of cross-product substitution effects. 

\newpage

\newpage
\subsection{Cereal products information}\label{cereal}
The following Table \ref{tab3} lists cereal products from FAOSTAT and their caloric content (based on FAO 2001) \cite{42}. Production data include only primary products, reserve data include rice, milled (code 31) and other primary products, and trade data include all products. Trade data for rice, paddy (rice milled equivalent) (code 30) was treated as rice, milled (code 31)\cite{26}. 

\newpage
\begin{table}[ht]
\centering
\footnotesize
\caption{List of cereal products.}\label{tab3}%
\begin{tabular}{@{}cp{5cm}cc@{}}
\toprule
Product code & Product name & Primary Product code & Conversion factor \\
 & & & ($10^6$ kcal/ton) \\
\midrule
\multicolumn{4}{@{}l@{}}{\textit{Wheat products}} \\
15 & Wheat & 15 & 3.34 \\
16 & Wheat and meslin flour & 15 & 3.64 \\
18 & Uncooked pasta, not stuffed or otherwise prepared & 15 & 3.67 \\
19 & Germ of wheat & 15 & 3.82 \\
20 & Bread & 15 & 2.49 \\
21 & Bulgur & 15 & 3.45 \\
22 & Pastry & 15 & 3.69 \\
41 & Breakfast cereals & 15 & 3.89 \\
\midrule
\multicolumn{4}{@{}l@{}}{\textit{Rice products}} \\
27 & Rice & 27 & 2.80 \\
28 & Husked rice & 27 & 3.57 \\
29 & Rice, milled (husked) & 27 & 3.57 \\
30 & Rice, paddy (rice milled equivalent) & 27 & 3.60 \\
31 & Rice, milled & 27 & 3.60 \\
32 & Rice, broken & 27 & 3.60 \\
38 & Flour of rice & 27 & 3.66 \\
\midrule
\multicolumn{4}{@{}l@{}}{\textit{Barley products}} \\
44 & Barley & 44 & 3.32 \\
45 & Pot barley & 44 & 3.48 \\
46 & Barley, pearled & 44 & 3.46 \\
48 & Barley flour and grits & 44 & 3.43 \\
49 & Malt, whether or not roasted & 44 & 3.68 \\
50 & Malt extract & 44 & 3.67 \\
\midrule
\multicolumn{4}{@{}l@{}}{\textit{Maize products}} \\
56 & Maize (corn) & 56 & 3.56 \\
57 & Germ of maize & 56 & 3.73 \\
58 & Flour of maize & 56 & 3.63 \\
\midrule
\multicolumn{4}{@{}l@{}}{\textit{Other cereal products}} \\
71 & Rye & 71 & 3.19 \\
72 & Flour of rye & 71 & 3.41 \\
75 & Oats & 75 & 3.85 \\
76 & Oats, rolled & 75 & 3.84 \\
79 & Millet & 79 & 3.40 \\
80 & Flour of millet & 79 & 3.40 \\
83 & Sorghum & 83 & 3.43 \\
84 & Flour of sorghum & 83 & 3.43 \\
\botrule
\end{tabular}
\end{table}

\newpage
\subsection{Country and region information}\label{country}
The following tables \ref{tab4}--\ref{tab8} list the 171 countries in the main text with their ISO3 codes and regional information. F15 (Belgium-Luxembourg) exists only in the data from 1993--1999, after which it is disaggregated into two separate country datasets: BEL (Belgium) and LUX (Luxembourg). SCG (Serbia and Montenegro) exists only in the data from 1993--2005, after which it is disaggregated into MNE (Montenegro) and SRB (Serbia). F206 (Sudan (former)) exists only in the data from 1993--2011, after which it is disaggregated into SDN (Sudan) and SSD (South Sudan).

\newpage
\begin{table}[ht]
\centering
\caption{List of 52 African countries with ISO3 codes.}\label{tab4}
\begin{tabular}{@{}ll|ll@{}}
\toprule
Country & ISO3 & Country & ISO3 \\
\midrule
Angola & AGO &  Lesotho & LSO \\
Burundi & BDI &  Morocco & MAR \\
Benin & BEN &  Madagascar & MDG \\
Burkina Faso & BFA & Mali & MLI \\
Botswana & BWA &  Mozambique & MOZ \\
Central African Republic & CAF &  Mauritania & MRT\\
Côte d'Ivoire & CIV &  Mauritius & MUS \\
Cameroon & CMR &  Malawi & MWI \\
Democratic Republic of the Congo & COD & Namibia & NAM \\
Congo & COG & Niger & NER\\
Comoros & COM & Nigeria & NGA \\
Cabo Verde & CPV & Rwanda & RWA \\
Djibouti & DJI & Sudan & SDN \\
Algeria & DZA & Senegal & SEN \\
Egypt & EGY &  Sierra Leone & SLE\\
Eritrea & ERI & Somalia & SOM \\
Ethiopia & ETH & South Sudan & SSD \\
Gabon & GAB & Eswatini & SWZ \\
Ghana & GHA & Chad & TCD \\
Guinea & GIN & Togo & TGO \\
Gambia & GMB & Tunisia & TUN \\
Guinea-Bissau & GNB & United Republic of Tanzania & TZA \\
Equatorial Guinea & GNQ & Uganda & UGA \\
Kenya & KEN & South Africa & ZAF \\
Liberia & LBR & Zambia & ZMB \\
Libya & LBY & Zimbabwe & ZWE \\
\botrule
\end{tabular}
\end{table}

\newpage
\begin{table}[ht]
\centering
\caption{List of 26 American countries with ISO3 codes.}\label{tab5}
\begin{tabular}{@{}ll|ll@{}}
\toprule
Country & ISO3 & Country & ISO3 \\
\midrule
Argentina & ARG & Haiti & HTI \\
Bolivia (Plurinational State of) & BOL & Jamaica & JAM \\
Brazil & BRA & Mexico & MEX \\
Canada & CAN & Nicaragua & NIC \\
Chile & CHL & Panama & PAN \\
Colombia & COL & Peru & PER \\
Costa Rica & CRI & Paraguay & PRY \\
Cuba & CUB & El Salvador & SLV \\
Dominican Republic & DOM & Suriname & SUR \\
Ecuador & ECU & Trinidad and Tobago & TTO \\
Guatemala & GTM & Uruguay & URY \\
Guyana & GUY &  United States of America & USA \\
Honduras & HND & Venezuela (Bolivarian Republic of) & VEN \\
\botrule
\end{tabular}
\end{table}

\newpage
\begin{table}[ht]
\centering
\caption{List of 47 Asian countries with ISO3 codes.}\label{tab6}
\begin{tabular}{@{}ll|ll@{}}
\toprule
Country & ISO3 & Country & ISO3 \\
\midrule
Afghanistan & AFG & Sri Lanka & LKA \\
United Arab Emirates & ARE & Maldives & MDV \\
Armenia & ARM & Myanmar & MMR \\
Azerbaijan & AZE & Mongolia & MNG \\
Bangladesh & BGD & Malaysia & MYS \\
Bahrain & BHR & Nepal & NPL \\
Bhutan & BTN & Oman & OMN \\
China, mainland & CHN & Pakistan & PAK \\
Cyprus & CYP & Philippines & PHL \\
Georgia & GEO & Democratic People's Republic of Korea & PRK \\
Indonesia & IDN & Palestine & PSE \\
India & IND & Qatar & QAT \\
Iran (Islamic Republic of) & IRN & Saudi Arabia & SAU \\
Iraq & IRQ & Singapore & SGP \\
Israel & ISR & Syrian Arab Republic & SYR \\
Jordan & JOR & Thailand & THA \\
Japan & JPN & Tajikistan & TJK\\
Kazakhstan & KAZ & Turkmenistan & TKM \\
Kyrgyzstan & KGZ & Timor-Leste & TLS \\
Cambodia & KHM & Türkiye & TUR \\
Republic of Korea & KOR & Uzbekistan & UZB \\
Kuwait & KWT & Viet Nam & VNM \\
Lao People's Democratic Republic & LAO & Yemen & YEM \\
Lebanon & LBN &  \\
\botrule
\end{tabular}
\end{table}

\newpage
\begin{table}[ht]
\centering
\caption{List of 38 European countries with ISO3 codes.}\label{tab7}
\begin{tabular}{@{}ll|ll@{}}
\toprule
Country & ISO3 & Country & ISO3 \\
\midrule
Albania & ALB & Italy & ITA \\
Austria & AUT & Lithuania & LTU \\
Belgium & BEL & Luxembourg & LUX \\
Bulgaria & BGR & Latvia & LVA \\
Bosnia and Herzegovina & BIH & Republic of Moldova & MDA \\
Belarus & BLR & North Macedonia & MKD \\
Switzerland & CHE & Malta & MLT \\
Czechia & CZE & Montenegro & MNE \\
Germany & DEU & Netherlands (Kingdom of the) & NLD \\
Denmark & DNK & Norway & NOR \\
Spain & ESP & Poland & POL \\
Estonia & EST & Portugal & PRT \\
Finland & FIN & Romania & ROU \\
France & FRA & Russian Federation & RUS \\
United Kingdom of Great Britain and Northern Ireland & GBR & Serbia & SRB \\
Greece & GRC & Slovakia & SVK \\
Croatia & HRV & Slovenia & SVN \\
Hungary & HUN & Sweden & SWE \\
Ireland & IRL & Ukraine & UKR \\
\botrule
\end{tabular}
\end{table}

\newpage
\begin{table}[ht]
\centering
\caption{List of 8 Oceania and not classified countries with ISO3 codes.}\label{tab8}
\begin{tabular}{@{}lll@{}}
\toprule
Region & Country & ISO3 \\
\midrule
Oceania & Australia & AUS \\
Oceania & Fiji & FJI \\
Oceania & New Zealand & NZL \\
Oceania & Papua New Guinea & PNG \\
Oceania & Solomon Islands & SLB \\
not classified & Belgium-Luxembourg & F15 \\
not classified & Sudan (former) & F206 \\
not classified & Serbia and Montenegro & SCG \\
\botrule
\end{tabular}
\end{table}

\end{document}